\documentclass[AMA,STIX1COL]{WileyNJD-v2}
\usepackage{moreverb}
\usepackage{epstopdf}
\usepackage{graphicx}
\usepackage{adjustbox,lipsum}
\usepackage{multirow}
\usepackage{pbox}
\usepackage{xcolor}
\newcommand\BibTeX{{\rmfamily B\kern-.05em \textsc{i\kern-.025em b}\kern-.08em
T\kern-.1667em\lower.7ex\hbox{E}\kern-.125emX}}

\articletype{RESEARCH ARTICLE}%

\received{<day> <Month>, <year>}
\revised{<day> <Month>, <year>}
\accepted{<day> <Month>, <year>}

\begin{document}
\title{SDSN@RT: a middleware environment for single-instance multi-tenant cloud applications}
\author[1,3]{Indika Kumara}
\author[2]{Jun Han}
\author[2]{Alan Colman}
\author[1,3]{Willem-Jan van den Heuvel}
\author[1,3]{Damian A. Tamburri}
\author[2]{Malinda Kapuruge}

\authormark{Indika Kumara \textsc{et al}}

\address[1]{\orgdiv{Tilburg School of Economics and Management}, \orgname{Tilburg University}, \orgaddress{\state{Tilburg, North Brabant, 5037 AB}, \country{Netherlands}}}

\address[2]{\orgdiv{School of Software and Electrical Engineering}, \orgname{Swinburne University of Technology}, \orgaddress{\state{Hawthorn, Victoria, 3122}, \country{Australia}}}

\address[3]{\orgdiv{Universiteit Jheronimus Academy of Data Science}, \orgaddress{\state{'s-Hertogenbosch, North Brabant, 5211 DA}, \country{Netherlands}}}

\corres{Indika Kumara, Tilburg University,Tilburg, North Brabant, 5037 AB,Netherlands. \email{i.p.k.weerasingha.dewage@tue.nl}}

\abstract[Abstract] {
With the Single-Instance Multi-Tenancy (SIMT) model for composite Software-as-a-Service (SaaS) applications, a single composite application instance can host multiple tenants, yielding the benefits of better service and resource utilization, and reduced operational cost for the SaaS provider. An SIMT application needs to share services and their aggregation (the application) among its tenants while supporting variations in the functional and performance requirements of the tenants. The SaaS provider requires a middleware environment that can deploy, enact and manage a designed SIMT application, to achieve the varied requirements of the different tenants in a controlled manner. This paper presents the SDSN@RT (Software-Defined Service Networks @ RunTime) middleware environment that can meet the aforementioned requirements. SDSN@RT represents an SIMT composite cloud application as a multi-tenant service network, where the same service network simultaneously hosts a set of virtual service networks (VSNs), one for each tenant. A service network connects a set of services, and coordinates the interactions between them. A VSN realizes the requirements for a specific tenant and can be deployed, configured, and logically isolated in the service network at runtime. SDSN@RT also supports the monitoring and runtime changes of the deployed multi-tenant service networks. We show the feasibility of SDSN@RT with a prototype implementation, and demonstrate its capabilities to host SIMT applications and support their changes with a case study. The performance study of the prototype implementation shows that the runtime capabilities of our middleware incur little overhead.}


\maketitle

\section{Introduction}
\noindent Computing has become a utility with the advancement of cloud computing. Cloud service models offer a wide range of resources, from computing resources to business services, as utilities for individual and organizational customers. In particular, with Software as a Service (SaaS) composite applications, the cloud service provider can share the use of resources, services and their aggregations among its customers to achieve economies of scale through multi-tenancy. 

\indent There are two main multi-tenancy models for composite cloud applications \cite{R1,R2}: multi-instance multi-tenancy (MIMT) and single-instance multi-tenancy (SIMT). In the MIMT model, a tenant uses a dedicated runtime application variant and associated services (see Figure 1(a)). In the SIMT model, the tenants simultaneously share the same runtime application instance and services (see Figure 1(b)). The separation in the MIMT model can simplify the support for tenant-specific variations and changes, while the sharing in the SIMT model can reduce operational cost and improve utilization of the capacities of services. An important goal for a multi-tenancy support is to achieve the best of these two multi-tenancy models at the same time, i.e., to support tenant-specific (functional and performance) variations and changes while achieving sharing in the SIMT model.
\begin{figure}[!b]
\centering
\includegraphics[scale=1,keepaspectratio]{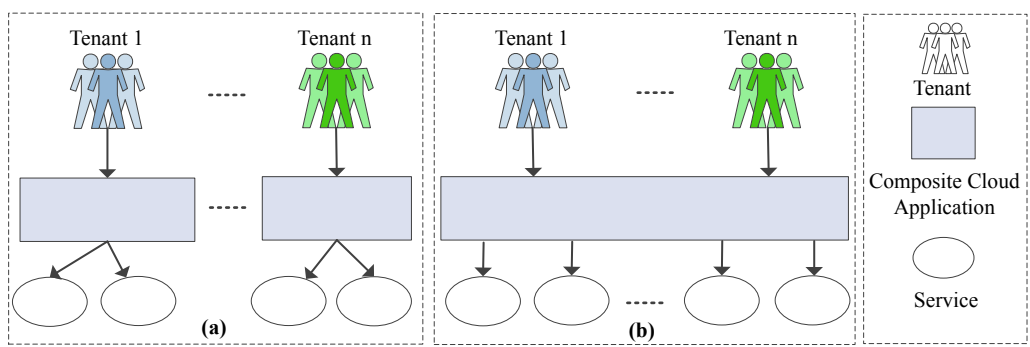}
\caption{Composite cloud applications: (a) MIMT model (b) SIMT model}
\end{figure}

\indent To host and manage composite SIMT cloud applications at runtime, a specialized middleware platform is needed. First, the middleware should create and maintain the runtime models of a composite application. These models need to connect the shared services according the relationships between the services,  and to coordinate the message exchanges between the services. Second, the middleware needs to be able to form and manage tenant-specific (\textit{virtual}) composite application variants over the shared services and application (i.e., slicing the shared services and application).  These variants need to reflect the commonality and variability in the functional and performance requirements of the tenants. The achievement of the performance requirements of a composite service application and its tenant-specific virtual variants depends on the performance of the used service capabilities, the architectural composition of these service capabilities (referred to as \textit{configuration design or topology}), and the regulation of the message exchanges among the relevant services (referred to as \textit{regulation design}) \cite{R3}. Thus, the middleware needs to support the sharing and variations in services, configuration designs, and regulation designs among application variants. Finally, the middleware should support the monitoring of, and runtime changes to, the shared cloud application. A software engineer or a management application needs to be able to use these middleware capabilities to monitor and modify the application. Note that, in this paper, we focus on \textit{business (IT) services}, which are surrogates for and support real world businesses of enterprises. The underlying businesses (rather than infrastructure-level IT resources) usually constrain the capacities of such services.

\indent While there are interesting works on middleware platforms for hosting general composite applications \cite{R4,R5,R6} and multi-tenant composite applications \cite{R7,R8,R9,R10,R11,R12}, they do not provide sufficient support to meet the above challenges posed by the SIMT model in achieving runtime sharing with variations. Existing middleware platforms for multi-tenant component compositions \cite{R7,R8} support the generation of application variants and the creation of runtime models for those variants (per a user request or per a tenant), and runtime binding of tenant-specific components to the declared locations in the component model of the application per a user request. Similarly,  middleware platforms for multi-tenant service compositions support the creation and separate deployment of service composition variants \cite{R9}, runtime binding of services per a user request \cite{R10}, variations through conditional branches and parameters \cite{R9, R10, R11}, and runtime changes to service compositions \cite{R12}. In general, the works that create the runtime models per application variants can only support the MIMT model. While the runtime binding or selection of tenant-specific variations (components, services, or code branches) at the specific locations in a shared application can support the SIMT model, it only provides a limited flexibility to form application variants by dynamically composing a subset of the elements in the application. For example, the fine-grained variants and variation points, and the tangling of normal processing logic with variant differentiation logic can make the application complex \cite{R13, R14}. Regarding the regulation design of multi-tenant applications, some works support tenant-specific policy enforcement \cite{R15}(without sharing or variation of policies among tenants), quality-aware application variant generation \cite{R16}, and specific \cite{R17,R18,R19,R20} or configurable \cite{R21} regulation mechanisms. However, these works have not considered the runtime sharing and tenant-specific variations in the regulation design or to runtime changes to the regulation design. 

\indent In this paper, we present \textit{SDSN@RT (Software-Defined Service Networks @ Run Time)}, which is a middleware platform that can enact and manage SIMT composite cloud applications that share services among tenant-specific (virtual) composite applications  while realizing the differing functional and performance requirements of the tenants. We represent such an SIMT application at runtime as \textit{a multi-tenant service network} or \textit{a software-defined service network (SDSN)}, where a set of virtual service networks (VSNs) simultaneously coexist on the same physical service network in a way analogous to virtual computer networks. The (physical) service network is a runtime structure that connects services, and mediates the interactions between the services. It is created based on the \textit{models@runtime} concept \cite{R22}. A VSN  belonging to a tenant meets the tenant's functional and performance requirements, and uses a subset of the services in the underlying physical service network. The different VSNs have similarities and differences in their services, configuration designs, and regulation designs. A VSN is enacted by consuming the service capabilities used by the VSN, and routing the relevant interaction messages (capability requests and responses) between relevant services in the service network. Through a separate controller, these interaction flows can be monitored and regulated according to the designs of the VSNs, and the multi-tenant service network can be modified and reconfigured at runtime.

\indent The main contribution of this paper is to address the above-mentioned requirements for a middleware platform that hosts SIMT composite cloud applications. To this end, we present 1) the runtime models of SIMT applications designed as multi-tenant service networks, 2) the support for simultaneous and controlled enactment of virtual application variants (i.e. VSNs), and 3) the support for policy-based runtime management of applications and their variants. We show the feasibility of SDSN@RT with a prototype implementation, demonstrate its capabilities with a case study, and quantify its runtime overhead in supporting the SIMT model.

\indent In an earlier paper \cite{R3}, we have presented the design support for SDSN (or the programming model of the SDSN@RT middleware) that allows a software engineer to define multi-tenant service networks using a set of design constructs and an associated domain-specific language. The paper also included a section (around 1 page) on the SDSN middleware to emphasize the fact that the designs are not merely conceptual models, but can be enacted/executed. However, the main focus of the paper was to describe our approach to supporting the functional and performance variability at the design specification level (design time variability). In contrast, this paper presents the SDSN middleware in detail, focusing on its support for maintaining the runtime models of the SIMT cloud application, achieving the runtime functional and performance variability via dynamic routing in service networks, and managing runtime changes to the application. In this paper, we only discuss the runtime abstractions in the application, and the architectural and runtime characteristics and capabilities of the middleware (in contrast, the focus of our earlier paper was design models).

\indent Our SDSN middleware have several key middleware novelties. First, the dynamic routing at the application architecture level is introduced as a flexible runtime variability mechanism. The application architecture (architectural elements and their relationships), and the conversational behaviors (message exchanges between the architectural elements) are made explicit, visible, and controllable. A set of abstractions are introduced to intercept, regulate, and route the message exchanges. Consequentially, the dynamic routing enables slicing the same runtime application at the architecture level, and dynamically controlling conversational behaviors in the application slices (as in flexible network-based models \cite{R23,R24}). Thus, the dynamic routing provides a higher controlled flexibility to achieve the runtime variability compared with the existing approaches such as selecting, activating, or injecting components/services within a fixed application structure (e.g., a component model). Moreover, the explicit architectural model of the composite application can improve the flexibility to dynamically manage/change the composite application at runtime (as in the applications with explicit configuration designs \cite{R25}). Second, to the best of the knowledge of the authors, our SDSN middleware is the only middleware to support the enactment of the composite service applications designed following the service network model, which is a service composition model that can provide a natural abstraction to design, enact, regulate, and manage webs of service-based businesses \cite{R26, R27}. Finally, our SDSN middleware uses the models@runtime, the dynamic routing, and the extended service network model to support the single instance multi-tenancy (SIMT) model for business service-based composite cloud applications.

\indent The rest of this paper is organized as follows. Section 2 presents a business example to motivate the SIMT model for composite cloud applications, and identifies the challenges for a middleware platform that hosts such applications. Section 3 presents our SDSN@RT middleware, including its system architecture, its runtime models for representing SIMT cloud applications, its capabilities to enact and manage such applications, and its prototype implementation. Section 4 presents the implementation of the two case studies with the SDSN@RT prototype, and includes a performance study of the prototype. Section 5 summarizes the related work, and  Section 6 concludes the paper, and outlines some future work. 
\section{Multi-tenant composite cloud applications and middleware challenges}
\noindent In this section, we first present a business-driven case study used in this paper, highlighting the multi-tenancy requirements of such business scenarios. Then, we describe the key challenges pertaining to a middleware that can enact and manage multi-tenant cloud applications supporting such business scenarios.
\subsection{Case study}
\noindent CampSAS is a company that offers camping assistance to its business clients (tenants) such as campsite owners and travel agencies (see Figure 2). Its camping assistance business is supported through a software application. This application adopts the SIMT SaaS model due to its resource utilization and operational benefits, and integrates partner business services such as camping equipment providers and tour companies. 

\indent The functional and performance requirements of the different tenants exhibit commonalities and variations. Figure 2 shows three tenants of CampSAS. They use tour arrangement and case handling capabilities (commonalities), and select two more capabilities from equipment rental, grocery delivery, and taxi hire (variations). Their performance requirements also vary. For example, the expected response time is 3 hours/case for the tenant HappyTours, and 4 hours/case for the tenant UniUvtClub.

\indent The partner services also have both similar and different capabilities, even for the same service type. For example, TomGear and JackGear are both camping equipment providers. Compared with TomGear, JackGear has an internal delivery service, and thus can deliver the rented equipments to the campers without using an external delivery company. Their rental capacities, which are the maximum number of rental orders per day, also differ (200 compared with 100). Note that these finite capacities of the business services cannot be changed by simply managing the computing resources the services use.
\begin{figure}
\centering
\includegraphics[width=\textwidth, height=\textheight, keepaspectratio]{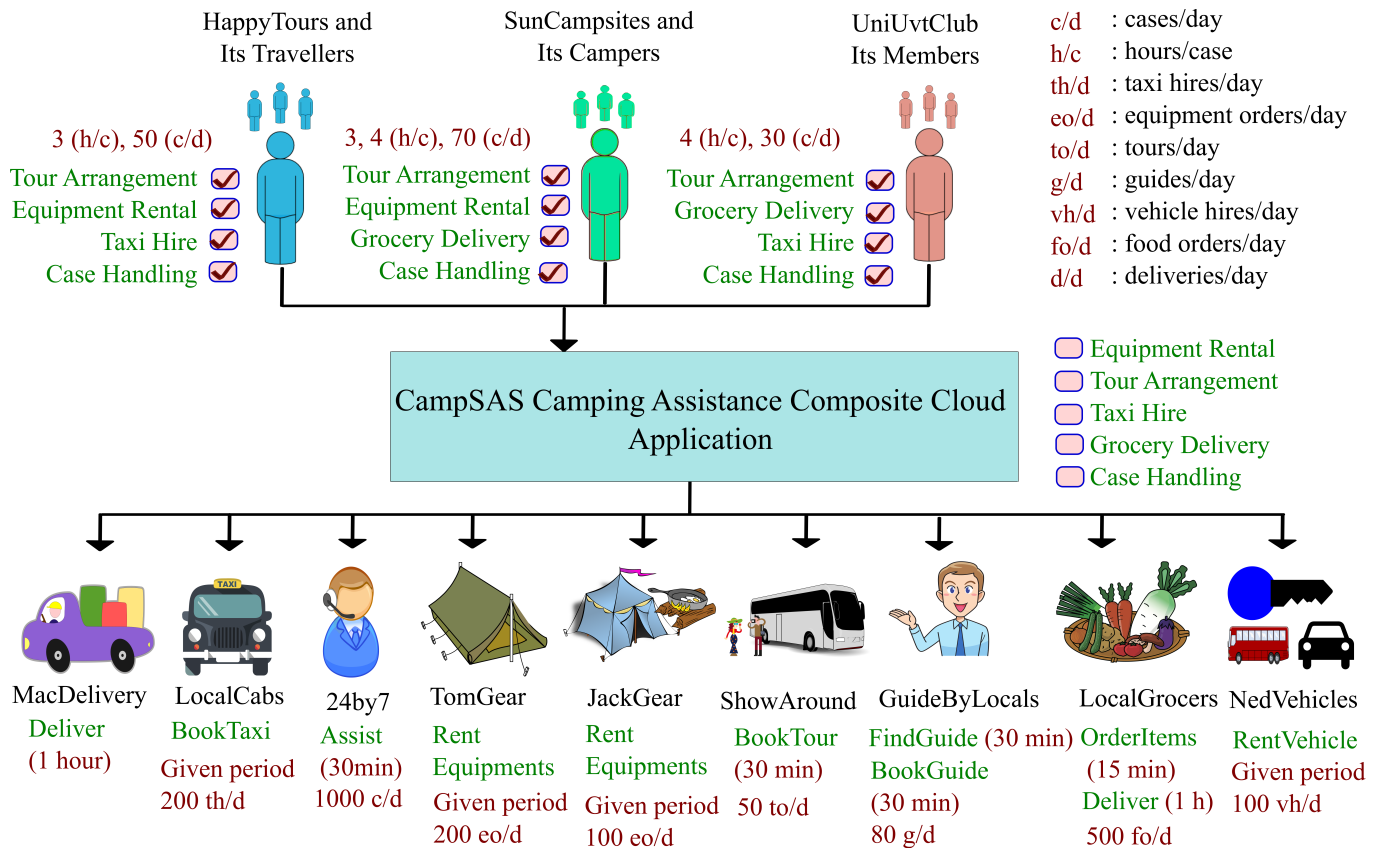}
\caption{CampSAS, its partner organizations, and its tenants and their customers}
\end{figure}
\subsection{Scenarios} 
The software engineers at CampSAS can design the camping assistance composite cloud application using a design methodology and tools for SIMT cloud applications, for example, our SDSN design support \cite{R3}. Once the application is designed, the software engineers need a middleware environment that can host the designed cloud application, and enable the enactment and runtime management of the deployed application. The following scenarios illustrate the usage of the middleware environment.  
\begin{itemize}
\item\textit {Deployment.} First, a software engineer deploys the camping assistance cloud application on the middleware by providing the design artifacts for the application. The deployed application connects the partner services to enable and manage the interactions between them. Second, a software engineer deploys the variants of the shared camping assistance application for the three tenants. These application variants share some subset of the partner services and the application, which reflects the commonality and variability in the functional and performance requirements of the tenants. 
\item\textit {Enactment and Monitoring.} The campers of the tenants request the assistance from CampSAS when they need some camping services. When an assistance request is received, the middleware creates and enacts an instance of the  camping assistance application variant of the corresponding tenant to support the requested camping assistance case. Multiple camping assistance cases coexist and progress simultaneously. The middleware monitors the progress of each camping assistance case and measures such metrics as the time to complete a case and the available capacity of the partner services. 
\item\textit {Management.} The requirements of tenants and CampSAS and the availability and capabilities of services can change over time. For example, after one month, the tenant SunCampsites decides that it needs 50 additional assistances per day. After two months, CampSAS decides to support the bike rental capability in its camping assistance business, and the tenant HappyTours would like to use the new bike rental capability instead of the currently used taxi hiring capability. To support these changes, a software engineer evolves the shared camping assistance cloud application by modifying its design at runtime without disturbing the operations of those tenants unaffected by the change (for example, the tenant UniUvtClub).
\end{itemize}
\subsection{Middleware challenges} 
The above scenarios represent three general key challenges for a middleware environment to support such SIMT composite cloud applications.
\begin{itemize}
\item\textit{Create and maintain the runtime models of the designed composite cloud application (Req1)}. To enable the enactment of and runtime incremental changes to the composite cloud application, the middleware needs to create and maintain suitable runtime models of the application \cite{R22,R28}. These models also need to facilitate the managed interactions between the services. 
\item\textit{Enact and control concurrent variants of the deployed composite cloud application (Req2)}. The middleware needs to create the variant instances of the composite cloud application for the instantiation requests from the customers of the tenants, and to progress the created instances by executing the elements of the runtime models of the application and the service operations that are required, and support the requirements of the tenant's application variant. These instances need to be able to share and selectively use the services and the runtime elements of the composite application as necessary to accommodate the similarities and differences in the functional and performance requirements of the tenants. The middleware also needs to enable fine-grained monitoring of the progress of the variant instances and the performance of the partner services. 
\item\textit{Enable the runtime management of the deployed composite cloud application and its variants (Req3).} The middleware needs to support different classes of runtime changes that can potentially occur to the SIMT applications. The changes need to be realized and managed at runtime without disturbing the operations of those tenants and end-users unaffected by the change. A software engineer should be able to declaratively specify changes, and enact the change specification at runtime to modify the application.  
\end{itemize}
\section{The SDSN@RT Middleware Environment} 
Software-Defined Service Networking (SDSN)\cite{R3} is an architecture-oriented approach for software engineers to design, enact, and manage SIMT service-based composite SaaS applications. SDSN represents a composite SaaS application as \textit{a multi-tenant service network}, where a set of managed virtual service networks (VSNs) share the same physical service network at runtime. A VSN represents a service composition in the service network for a particular tenant. The concept of the multi-tenant service network realizes the SIMT model. In contrast, physically separated service networks realize the MIMT model. In the introduction section, we have compared the SIMT model with the MIMT model.

\indent This section presents SDSN@RT, which is a middleware runtime that supports the deployment, enactment, monitoring, and management of SIMT composite cloud applications designed as multi-tenant service networks following our SDSN approach. We first provide an overview of the SDSN@RT system architecture (Section 3.1). Upon the deployment of the design models of a multi-tenant service network, their runtime models are created following the \textit{models@runtime} concept \cite{R22}. The runtime model of a service network acts as the \textit{network infrastructure} that allows the controlled interactions (exchange of messages) between services according to the design, and thus supports the enactment of virtual service networks (VSNs). A service network can be monitored and modified at runtime. Sections 3.2 and 3.3 describe the structure and the responsibilities of each runtime abstraction in multi-tenant service networks. Section 3.4 discusses the enactment of a VSN, while Section 3.5 presents the management capabilities of SDSN@RT. In Section 3.6, we briefly describe the prototype implementation of SDSN@RT.
\subsection{Overview of SDSN@RT system architecture}
\noindent Figure 3 shows the high-level system architecture of SDSN@RT. The architecture has three main layers: services, coordination infrastructure, and management platform. 
\begin{figure}[!t]
\centering
\includegraphics[scale=1.2,keepaspectratio]{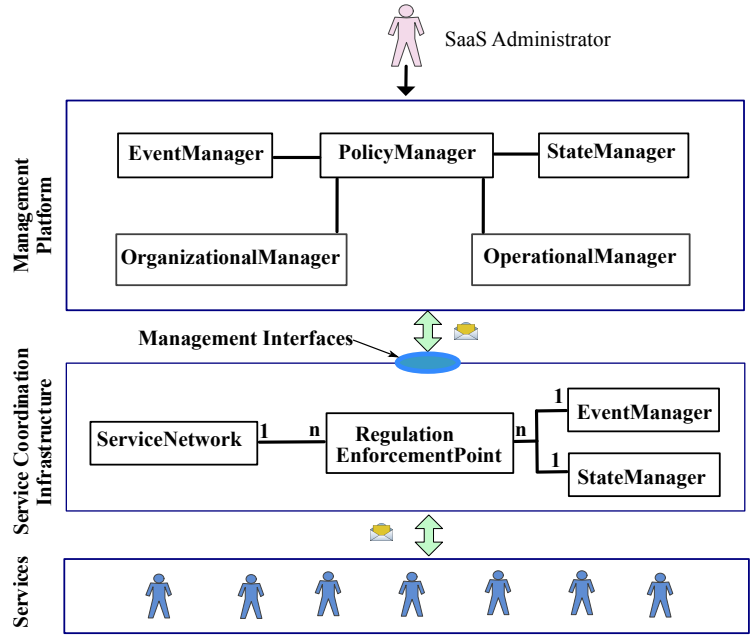}
\caption{High-level system architecture of SDSN@RT (with a single deployed service network)}
\end{figure}

The services are Web services that support and act as proxies to the real world business capabilities of the enterprises. Multiple tenants in a business service network share the services and their capacities. These services contractually relate to each other and collaborate at runtime by exchanging messages. These interaction messages between services need to be intercepted, coordinated and regulated to achieve the functional and performance goals of the SaaS provider and its tenants. For this purpose, the service coordination infrastructure creates and maintains a runtime model of the service network (i.e., an overlay network over the services). A service network (runtime model) has a set of Regulation Enforcement Points (REPs) to allow the observation and regulation of the message flows throughout the service network. The coordination infrastructure also has an event manager and a state manager. The event manager records the events being generated in the service network, for example, service interaction events and performance violation events. The state manager records the states of the service network, such as the used service throughput and average response time. The events and states can trigger the regulation policies at REPs, and thus regulate the enactment of VSNs. The event manager and the state manager support the publish-subscribe model for publishing and consuming events and state records, respectively. 

\indent The management logic of a service network resides mainly in the management platform, which acts as \textit{the brain of the service network}. For a given multi-tenant service network, the management platform consists of an organizer and an operational manager that enforce configuration and regulation management policies, respectively. To implement their capabilities, both managers use a policy manager, an event manager, and a state manager. The policy manager stores, and executes the management policies. The behaviors of the event manager and the state manager are similar to those at the coordination infrastructure, except they control the enactment of the management policies. 

\indent The management interfaces between the management platform and the coordination infrastructure define the allowed management interactions between them. There are three types of management interfaces: deployment, organizational management, and operational management. The deployment interface is at the system level, and supports the deployment and un-deployment of (multi-tenant) service networks. The latter two interfaces are at the application level (per each service network), and support, respectively, the management of the configuration and regulation designs of a running multi-tenant service network at runtime. The organizer and operational manager of the service network can use these management interfaces to dynamically change the topology of the service network as well as the routing and regulation of the messages passing over the service network.

\indent The management platform also offers its management capabilities as Web services interfaces so that the SaaS administrator can remotely deploy the designed services networks, monitor and modify the deployed service networks by providing management policies, and un-deploy the deployed service networks as necessary.

\indent SDSN@RT can enact multiple multi-tenant service networks. Each deployed service network has its separate and isolated instances (Java objects) of the architectural elements such as service network, REP, event manager, state manager, policy manager, organizational manager, and operational manager, as the runtime representations of the corresponding virtual service network instances. In other words, each deployed service network has a separate runtime model (i.e., a Java object model). For simplicity, we only consider a single service network in this paper.
\subsection{Service networks}
\noindent A composite SIMT SaaS application relies on a range of services for supporting its business. To represent, compose and manage these services in the context of the entire application, \textit{roles} are introduced to represent the services, and \textit{contracts} between services to connect the services for their interactions. The network of roles (\textit{nodes}) and contracts (\textit{links}) forms the \textit{configuration} or structure of the network, and is superimposed on the services for their management and interaction. The service network acts as an overlay network over the services. A role comprises a set of \textit{tasks}, which represent the used capabilities of the service. It acts as a proxy service to the external service, and can send (Web service) messages to the service and receive messages from the service. A contract consists of a set of \textit{interaction terms}, and the references to the two roles that it connects. The tasks and interaction terms define the message templates, which are used to create and validate the role-role and role-service interaction messages.
\begin{figure}[!t]
\centering
\includegraphics [scale=0.4,keepaspectratio]{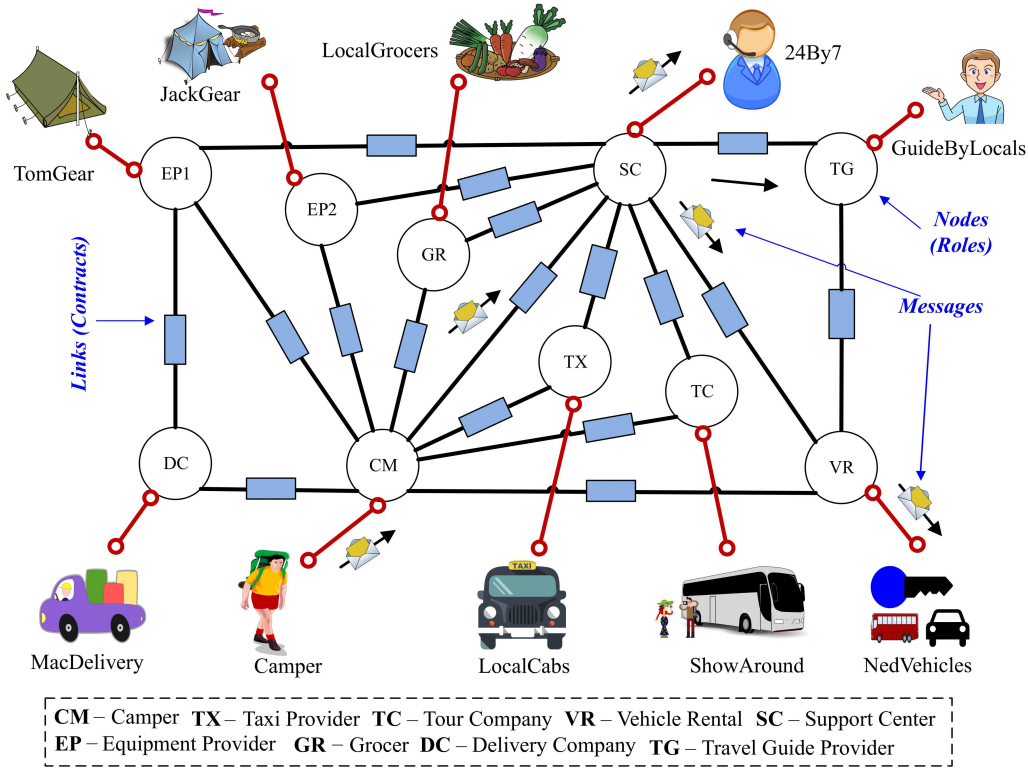}
\caption{The camping assistance service network of CampSAS}
\end{figure}

\indent Figure 4 shows the service network for our motivating example. It consists of a number of roles (e.g., CM, SC, EP1, and DC) connected by contracts (e.g., CM\_SC, CM\_DC, CM\_EP1, and SC\_EP1) to form the service network supporting the coordination of the interactions between the services (e.g., 24by7 support center, TomGear equipment provider, and MacDelivery delivery service) to meet the camping assistance requirements of the tenants. As shown in Figure 5, the roles include the relevant tasks, for example, \textit{tRentEquipment} of the role EP1 (to rent camping equipments) and \textit{tDeliver} of the role DC (to deliver the rented equipment to the campers). The contracts include the relevant interaction terms, for example, \textit{iSendEquipmentRequirements} of the contract CM\_EP1 (to represent sending the equipment requirements to the rental company) and \textit{iOrderEquipment} of the contract SC\_EP1 (to represent the approval from the support center for renting the required equipments).

\indent The \textit{regulation} of the message flow over the service network structure is achieved through a set of regulation enforcement points (REPs) at the various roles and contracts of the network. The messages flow from REP to REP, which can intercept messages and apply various operations on the intercepted messages. There are four types of REPs: synchronization (at each role), routing (at each role), pass-through (at each contract), and coordinated-pass-through (across contracts). The synchronization REP of a role synchronizes a subset of incoming interactions from the adjacent roles (and consequently their services) before invoking a task of the service represented by the role by sending a request message to the service. The routing REP of a role receives and routes a response or request message from the role's service to a subset of the adjacent roles (and consequently their services). The pass-through REP can intercept and process the interaction messages between two roles (and thus their services). The coordinated-pass-through is to regulate the interactions across different pairs of roles (and their services).

\indent Each REP consists of a \textit{regulation knowledgebase} and a \textit{regulation table}. The knowledgebase contains a set of event-condition-action (ECA) rules that define the regulation logic. The regulation table maps a service network flow to one or more ECA rules, which determine the treatment of the flow. We define a service network flow broadly as a set of messages belonging to a specific entity such as a virtual service network (VSN), a business process, and a VSN instance (for an end-user). The messages and their interpretations (i.e., events and recorded states) carry the flow information (e.g., the identifiers of the relevant entities). This flow information allows the enactment engine to isolate the executions of a given composite application by the end-users of different tenants (similar to the use of the tenant-context in \cite{R9} for tenant execution isolation).

\indent The condition part of a regulation rule can be a logical expression of message contents, events, and state records. Thus, the arrival of a message at a REP, and a new or changed service network event or state can trigger a subset of rules. The action part of a rule uses one or more regulation functions, which tell the REP what to do with the individual interaction messages and/or the service network flow. A regulation function implements a unit of regulation decisions on messages such as admission control, transformation, service operation execution,  response time and throughput monitoring, and event publishing.  A detailed discussion of the available regulation functions can be found in our previous publication \cite{R3}. They collectively support the connection and interaction between services, and the monitoring and regulation of service interactions. 
\begin{figure}[!b]
\centering
\includegraphics [scale=0.5,keepaspectratio]{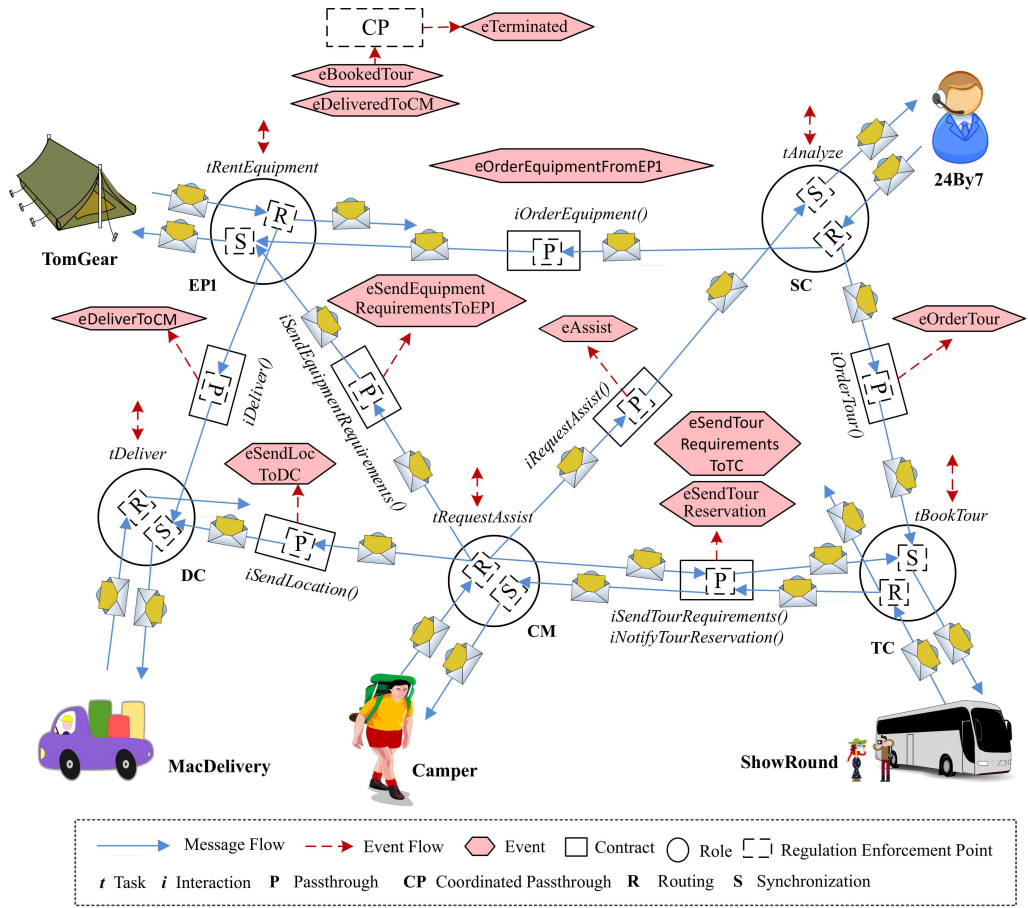}
\caption{A part of the regulation design of CampSAS's service network}
\end{figure}
\indent There may be control flow relations between regulation functions as well as between rules. The underlying rule engine determines these control flows and the activation of the individual rules. For this purpose, we use a Rete algorithm \cite{R29} based rule engine, namely Drools (drools.org). The execution of the rules for different service network flows is independent and isolated.

\indent Figure 5 illustrates the four types of REPs in a part of the camping assistance service network. Each role in the CampSAS service network has a routing REP and a synchronization REP, and each contract has a pass-through REP. The CampSAS service network as a whole also has a coordinated pass-through REP. Figure 6(a) shows three rules at the routing REP of the role. These rules control the admission of the assistance requests, and initiate the send equipment requirements and send location role-role interactions towards roles EP1 and DC, respectively. Figure 6(b) shows a pass-through rule at the REP of the contract CM\_DC, which evaluates the interaction message \textit{iSendLocation}, and generates the event \textit{eSendLocToDC} to indicate the occurrence of the interaction. Figures 6(c) and 6(d) show two synchronization rules, one at the role EP1, and other at the role DC. The first rule ensures that the equipments are rented based on the requirements provided by the camper, upon receiving the rental order from the 24by7 support center. The second rule ensures that MacDelivery service delivers the rented camping equipments to the location given by the camper. Figure 6(e) shows a regulation rule at the coordinated pass-through REP. It adjusts the used throughput for a virtual service network as its running instances end. The termination of a VSN instance is identified by the occurrence of the interaction \textit{iDeliver} (response) between roles EP1 and DC (indicated by the event \textit{eDeliveredToCM}), the interaction \textit{iSendTourReservation} (response) between roles TC and CM (indicated by the event \textit{eTourReservationSent}), and the interaction \textit{iSendTaxiReservation} (response) between roles TX and CM (indicated by the event \textit{eTaxiReservationSent}).
\indent 
\begin{figure}
\centering
\includegraphics [scale=0.55,keepaspectratio]{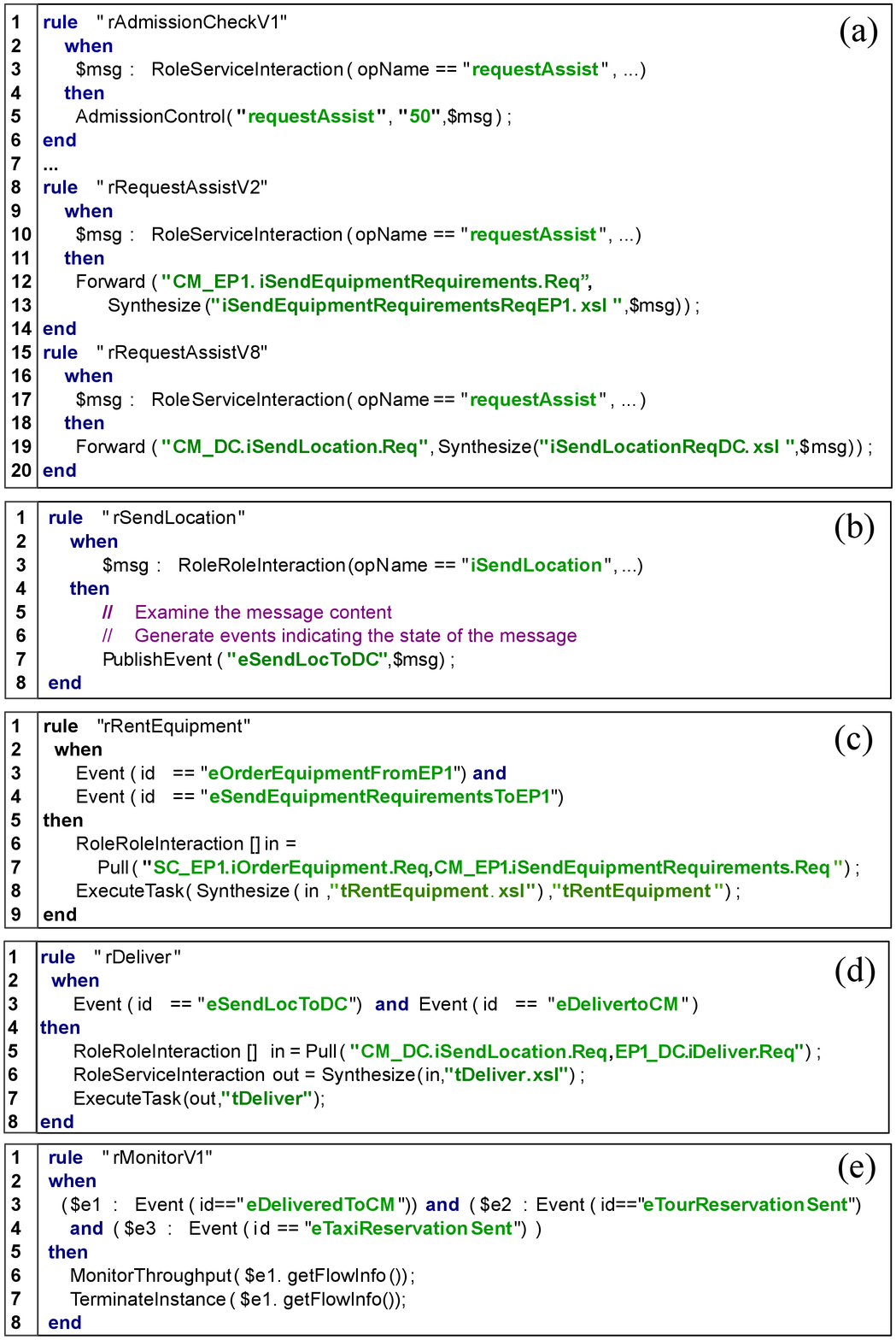}
\caption{Some regulation rules at: a) the routing REP of the role CM, b) the pass-through REP of the contract CM\_DC, c) the synchronization REP of the role EP1, d) the synchronization REP of the role DC, and e) the coordinated pass-through REP}
\end{figure}
\subsection{Virtual service networks}
To support the functional and performance requirements of a tenant, a virtual service network (VSN) for the tenant is formed on the service network. A VSN represents one or more service compositions or business processes in the service network. Each such process has a specific service network path/topology. To enable and control the routing of the interaction messages over its service network paths, a VSN has a regulation policy, which refers to a subset of regulation rules at the REPs on the relevant paths. The virtualization of the service network is achieved by isolating the messages belonging to each VSN/process, and routing them on the path of the VSN/process, while respecting the relevant regulation policy. 

\indent Multiple VSNs (of tenants) share some service network elements for their common requirements, and use some other service network elements for their distinctive requirements. The service network elements include roles, contracts, tasks, interaction terms, REPs and their regulation rules, partner services and their capabilities and throughput/capacities. 

\indent Figure 7 shows the service network topologies of the VSNs for the tenants HappyTours and UniUvtClub in our motivating scenario. The overlaps between the topologies show the sharing between the two VSNs. To support the similar requirements of taxi hiring and case handling, the two VSNs share the services 24by7Support and LocalCabs, and the elements in the service network that correspond to the service collaborations that support these requirements. The sharing of a service task also implies that of its throughput/capacity. To support the distinctive requirements of equipment rental and grocery delivery, the CampSAS's VSNs use different services and elements in the service network. HappyTours uses TomGear and MacDelivery, while UniUvtClub uses LocalGrocers. While both HappyTours and UniUvtClub need the tour arrangement capability, the capacity of the tour company ShowAround is limited. Moreover, the overall performance requirements of the two tenants (3 hours and 4 hours) are different. Thus, the two VSNs use different service collaborations for realizing the tour arrangement. A collaboration involving a vehicle rental company and a travel guide provider implements the tour arrangement for the tenant UniUvtClub.      
\begin{figure}
\centering
\includegraphics [scale=0.45,keepaspectratio]{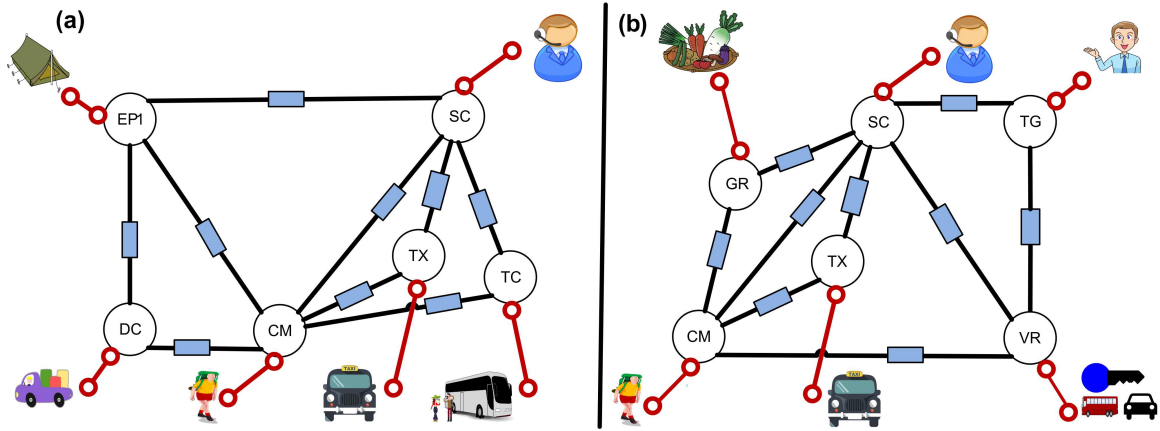}
\caption{The service network topologies of two VSNs in CampSAS' service network a) HappyTours, b) UniUvtClub}
\end{figure}

\indent The business processes in VSNs can be described and enacted as EPC (event-driven process chain) processes \cite{R30}. Events are used to define the temporal dependencies between the tasks used by processes. The regulation rules that produce and/or consume events support the enactment of the processes. Figure 8 shows a part of a process in the VSN of the tenant HappyTours. In parallel, an assistance request is submitted to the service 24by7, the location details are provided to the services MacDelivery and LocalCabs, the equipment requirements are sent to the service TomGear, and the tour requirements are sent to the service ShowAround. The events \textit{eAssist}, \textit{eSendLocToDC}, \textit{eSendLocToTX}, \textit{eSendEquipmentRequirementsToEP1}, and \textit{eSendTourRequirementsToTC} indicate these steps. 24by7 analyzes the assistance request, creates a support case, and approves the requested camping services (renting equipments, hiring a taxi, and booking a tour). These process steps are indicated by the task \textit{tAnalyze} and the relevant following events such as \textit{eOrderTour} and \textit{eOrderTaxi}. These events trigger some more tasks, and the completion of those tasks generate more events. For example, the events \textit{eOrderTaxi} and \textit{eSendLocToTX} trigger the task \textit{Tx.tBookTaxi}, whose completion generates the events \textit{eSendTaxiReservation} and \textit{eNotifyTaxiStatus}.
\begin{figure}
\centering
\includegraphics [scale=0.62,keepaspectratio]{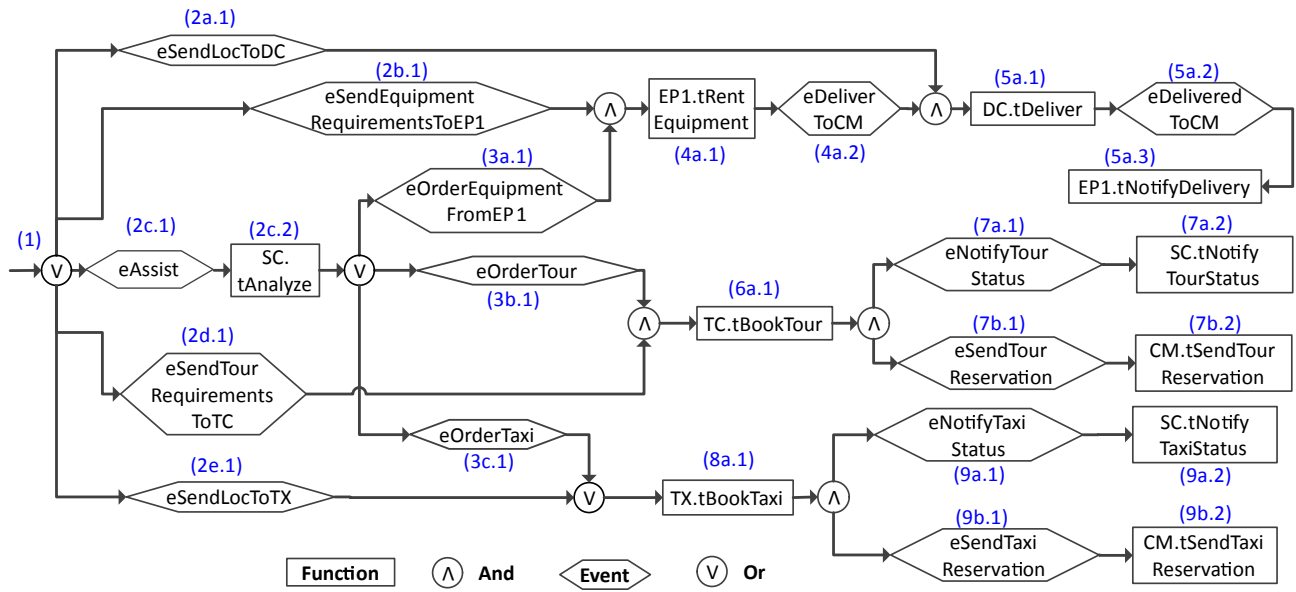}
\caption{An EPC diagram showing a part of the business process in HappyTours's VSN}
\end{figure}

\indent In the context of this paper, software engineers design the VSNs. They decide the service network paths for VSNs and estimate their performance to check their ability to satisfy tenants' functional and performance requirements. They also decide the regulation policies for VSNs to ensure that each instance of a given VSN/process follows the path of the VSN/process, and achieves the desired performance targets at runtime. Interested reader are referred to \cite{R3} for more details on the designs of VSNs, their processes, and their sharing of the service network with appropriate functional and performance variations.
\subsection{Enactment of VSNs as controlled message routing in service network paths}
To consume a business process of a tenant, an end-user of the tenant needs to send a Web service request to the service network via its user role, which represents the end-user (the role CM in our example). An instance of a business process in the tenant's VSN is created, and the created instance progresses by executing the relevant tasks (and consuming the relevant service capabilities) while respecting the control flow between the tasks. This creates interactions between services. These interaction messages are isolated from those of other process instances using the process instance identifiers included in the messages (as message headers). The isolated messages routed in the service network path used by the business process. Below, we discuss the enactment of the business processes in VSNs in detail. 

\indent In the first stage, when the routing REP of the user role receives an instantiation Web service request, the routing REP locates the regulation rules for the VSN via the regulation table. The Web service request contains the identifier of the VSN encoded in a message header or the part of the Web service endpoint reference of the user role (similar to the use of the tenant identifier in \cite{R9}). The VSN identifier is the key to the regulation table. The selected rules are executed by providing the received message as inputs. The consequences of the execution of the rules depend on the regulation functions used by the rules and the execution order of the functions. In general, the admittance of an instantiation request (\textit{AdmissionControl} function) is decided, and a business process for the admitted request is selected from the processes of the VSN to balance the load among processes (\textit{LoadBalance} function). Next, an instance of the process is created and maintained. A process instance represents an execution of a business process for a user request. It is a state object (non-executable) that records the execution states of the tasks in the process. Afterwards, the instantiation request is treated as a role-service interaction message, and is augmented with process and process instance identifiers and re-injected into the routing REP of the user role. If an instantiation request cannot be admitted at a given moment, it can be dropped or queued by using suitable regulation functions (\textit{Drop} and \textit{Schedule}). 

\indent In the second stage, the created process instance progresses as an EPC process. On receiving a role-service interaction message belonging to a process instance, the routing REP of the role applies the regulation rules for the corresponding process. In general, the regulation rules may control the admittance of the interaction message, select a set of role-role interactions according to the service network path of the process instance, convert the service interaction into role-role interaction messages, and schedule the created messages to be sent to the appropriate destination roles via the pass-through REPs at the relevant contracts. When the pass-through REPs receives the role-role interaction messages, the pass-through rules belonging to the process are applied to them. These rules may examine the interaction messages, generate events indicating the states of the messages, and forward the messages to the synchronization REPs at the destination roles. When the synchronization REPs receives the role-role interaction messages, they store the messages as the execution of task may need to synchronize multiple such interaction messages. In general, the regulation rules at the synchronization REPs use a logical expression of  events (generated at the pass-through REPs) as their triggering conditions. These rules retrieve the stored role-role interaction messages, create a role-service interaction  message from the retrieved messages, and send the created message to the destination service to carry out the desired task and consume the service capability. 

\indent A service may send messages via its role to indicate the completion of the requested capability or to request new capabilities from other services. These messages from the services will trigger new cycles of routing, pass-through, synchronization, and task execution. These regulated message exchanges between services in the service network path of a process continue until the process instance is terminated, which is the final stage of the VSN enactment. A process instance may terminate naturally when all its tasks are completed or when a certain condition is met (e.g., the completion of one or more specific tasks). In general, the regulation rules at the coordinated pass-through REP control the state of a process instance as multiple interactions between different pairs of services may need to be considered to decide the state of a process instance. Alternatively, the administrator can use the management capabilities provided by the service network to explicitly end a process instance.
\begin{figure}
\centering
\includegraphics [scale=0.75,keepaspectratio]{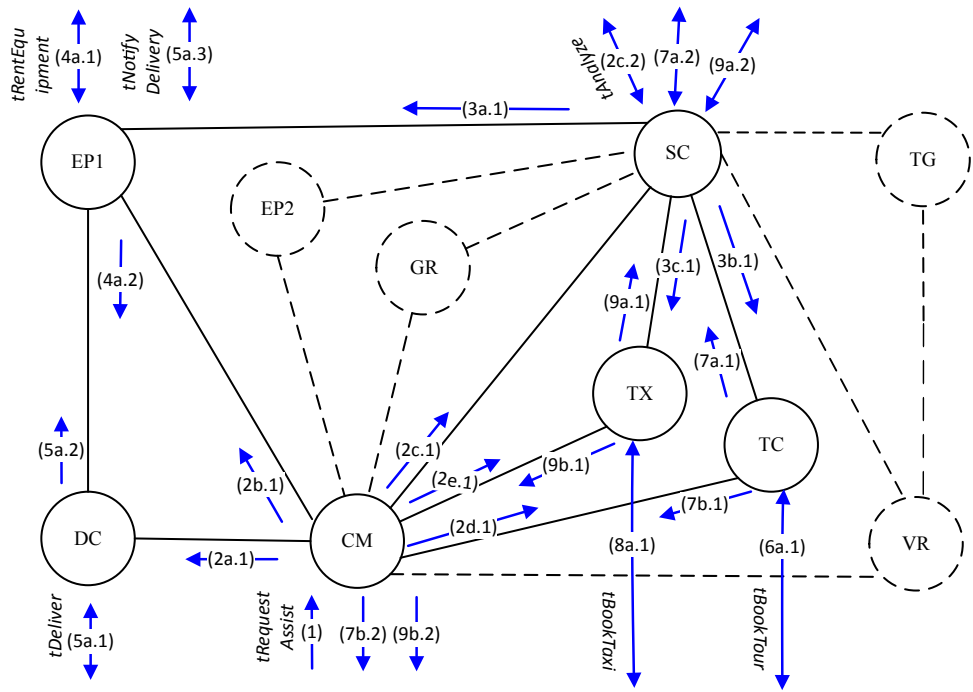}
\caption{The illustration of message flows in the service network when enacting the business process shown in Figure 8}
\end{figure}

\indent Figure 9 illustrates the message flow over the service network for the business process shown in Figure 8. The annotated numbers in both figures show the correspondence between the progress of business process and the routing of messages. For example, the request assist message from the client is routed (at the role CM) to the roles DC, EP1, SC, TC, and TX (2a.1, 2b.1, 2c.1, 2d.1, and 2e.1 in Figure 9) via the relevant contracts (pass-through REPs). This generates the events \textit{eSendLocToDC}, \textit{eSendEquipmentRequirementsToEP1}, \textit{eAssist}, \textit{eSendTourRequirementsToTC}, and \textit{eSendLocToTX} (2a.1, 2b.1, 2c.1, 2d.1, and 2e.1 in Figure 8). The generated events trigger some tasks (e.g., the event \textit{eAssist} triggers the task \textit{tAnalyze} of the role SC), which initiates the message flows from the relevant roles to their services (e.g., 2c.2 in Figure 9). The response messages from the services are routed to the relevant roles (and their services) via the relevant contracts, which generates more events. For example, the response message for the task \textit{tAnalyze} is routed to the roles EP1, TC, and TX (3a.1, 3b.1, and 3c.1 in Figures 8 and 9). The message routing in the service network path of the virtual service network (or its business process) continues until the business process is completed. 
\subsection{Runtime management of multi-tenant service networks}
The three common management tasks for an SIMT multi-tenant application are 1) the deployment and un-deployment of the application, 2) the deployment, un-deployment, and reconfiguration of the (virtual) application variants for tenants, and 3) the monitoring of and runtime modifications to the application and its variants. SDSN@RT supports these management tasks for multi-tenant service networks. The software engineers can use the deployment interfaces of the SDSN@RT to deploy or undeploy multi-tenant service networks. Via the organizational and operational management interfaces of the service networks, the software engineers can deploy and undeploy VSNs, modify deployed service networks, and monitor the business processes in VSNs. These management decisions can be specified as ECA rules.  Each management interface is exposed as a Web service interface. 
\begin{table}[!b]
\centering
\caption{Organizational management API}
\begin{adjustbox}{max width=\textwidth}
\begin{tabular}{|l|l|}
\hline
\textbf{Operation Name} & \textbf{Explanation} \\ 
\hline 
addRole (rId, rName, sBinding) & Add a new role (node) to the service network.\\ 
\hline
removeRole (rId) & Remove a role from the service network\\ 
\hline
updateRole (rId, property, value) & Update a property of a role\\ 
\hline
addTask (rId, tId, inputs, outputs) & Add a new task to a role\\ 
\hline
removeTask (rId, tId) & Remove a task from a role\\ 
\hline
updateTask (rId, tId, property, value) & Update a property of a task\\ 
\hline
addContract (cId, roleA, roleB) & Add a new contract (link) between two roles\\ 
\hline
removeContract (cId) & Remove a contract from the service network \\ 
\hline
updateContract (cId, property, value) & Update a property of a contract \\ 
\hline
addTerm (cId, tmId, direction) & Add a new interaction term to a contract \\ 
\hline
removeTerm (cId, tmId) & Remove an interaction term from a contract \\ 
\hline
updateTerm (cId, tmId, property, value) & Update a property of an interaction term \\ 
\hline
\end{tabular}
\end{adjustbox}
\caption*{\textit{rId}= role Id \textit{rName} = role name \textit{sBinding}=service binding \textit{tId} = task Id \textit{cId} = contract Id \textit{tmId} = term Id. The \textit{inputs} and \textit{outputs} of a task are references to interaction terms consumed/produced by the task. The \textit{direction} of an interaction term indicates the direction of the message flow between the relevant two roles.}
\end{table} 
\begin{table}[!b]
\centering
\caption{Operational management API}
\begin{adjustbox}{max width=\textwidth}
\begin{tabular}{|l|l|} 
\hline
\textbf{Operation Name} & \textbf{Explanation}  \\ 
\hline 
addSynchronizationRules (rId, ruleFile)  \textit{rId}= role Id& Add some rules to a synchronization REP\\ 
\hline
removeSynchronizationRules (rId, ruleIds) & Remove some rules from a synchronization REP\\ 
\hline
updateSynchronizationRule (rId, ruleId, key, value)  & Update a synchronization rule\\ 
\hline
addRoutingRules (rId, ruleFile) & Add some rules to a routing REP\\ 
\hline
removeRoutingRules (rId, ruleIds) & Remove some rules from a routing REP\\ 
\hline
updateRoutingRule (rId, ruleId, key, value) & Update a routing rule\\ 
\hline
addPassthroughRules (cId, ruleFile) \textit{cId} = contract Id & Add some rules to a pass-through REP\\ 
\hline
removePassthroughRules (cId, ruleIds) & Remove some rules from a pass-through  REP\\ 
\hline
updatePassthroughRule (ruleId, key, value) & Update a pass-through rule\\ 
\hline
addCoordinatedPassthroughRules (ruleFile) & Add some rules to a coordinated pass-through REP\\ 
\hline
removeCoordinatedPassthroughRules (ruleIds) & Remove some rules from a coordinated pass-through  REP\\ 
\hline
updateCoordinatedPassthroughRule (ruleId, key, value) & Update a coordinated pass-through rule \\ 
\hline
addSynchronizationTableEntries (snfId, entries) & Add some entries to the regulation tables at the \\
\textit{snfId} = service network flow Id & synchronization REPs on a path of a service network flow \\
\hline
removeSynchronizationTableEntries (snfId, entries) & Remove some entries from the regulation tables at the \\ & synchronization REPs on a path of a service network flow\\ 
\hline
updateSynchronizationTableEntry (snfId, & Update a synchronization regulation \\ entry, key, value) & table entry belonging to a service network flow \\ 
\hline
addRoutingTableEntries (snfId, entries) & Add some entries to the regulation tables at the \\ & routing REPs on a path of a service network flow \\ 
\hline
removeRoutingTableEntries (snfId, entries) & Remove some entries from the regulation tables at the \\ & routing REPs on a path of a service network flow\\ 
\hline
updateRoutingTableEntry (snfId, entry, key, value) & Update a routing regulation table entry belonging to a \\ & service network flow\\ 
\hline
addPassthroughTableEntries (snfId, entries) & Add some entries to the regulation tables at the \\ & pass-through REPs on a path of a service network flow\\ 
\hline
removePassthroughTableEntries (snfId, entries) & Remove some entries from the regulation tables at the \\ & pass-through REPs on a path of a service network flow\\ 
\hline
updatePassthroughTableEntry (snfId, entry, key, value) & Update a pass-through regulation table entry \\ &  belonging to a service network flow\\ 
\hline
addCoordinatedPassthroughTableEntries (snfId, entries) & Add some entries to the regulation table (belonging to a\\ &service network flow) at the coordinated pass-through REP \\ 
\hline
removeCoordinatedPassthroughTableEntries (snfId, & Remove some entries from the regulation table at\\ entries) & the coordinated pass-through REP\\ 
\hline
updateCoordinatedPassthroughTableEntry (snfId,& Update a coordinated pass-through regulation table\\ entry, key, value) & entry belonging to a service network flow\\ 
\hline
subscribe (eventPattern, subscriberEndpoint) & subscribe to a set of service network events\\ 
\hline
unsubscribe (eventPattern, subscriberEndpoint) & unsubscribe from a set of service network events\\ 
\hline
subscribe (stateIds, subscriberEndpoint) & subscribe to a set of service network states\\ 
\hline
unsubscribe (stateIds, subscriberEndpoint) & unsubscribe from a set of service network states\\ 
\hline
\end{tabular}
\end{adjustbox}
\end{table}
\begin{itemize}
\item \textit{Deployment}. The deployment operation takes as input the design artifacts of a service network, which include the descriptions of the roles, contracts, the regulation rule files, and the artifacts used by the regulation functions such as message transformation files. The un-deployment operation takes as input the name of the service network to be removed. Each deployed service network has an organizational manager and an operational manager in the management platform. These managers in turn use the organizational and operational interfaces offered by the deployed service network (in the service coordination infrastructure).  
\item \textit{Organizational management}. These management capabilities support the changes to the service network topology, which include addition/removal/update of roles, contracts, tasks, and interaction terms. Table 1 shows the main operations of the organizational management API.
\item \textit{Operational management}. These management capabilities allow the changes to the regulation design of a service network, which include addition/removal/update of regulation rules, regulation table entries, and service network events/states. REPs are created or removed as part of the creation or removal of their place-holder elements (i.e., roles, contracts, and service network). A process instance can also be paused, resumed, and terminated (by changing its management state). To monitor process instances, the service network events and states can be read by querying them or by subscribing to them. Table 2 shows the main operations of the operational management API. The parameter \textit{key} in the update operations indicate what to update, for example, management state or regulation rule content. A service network flow can represent the messages belonging to a VSN, a process in a VSN, or a process instance.
\end{itemize}
\indent To support the runtime changes to a multi-tenant service network in a controlled manner without compromising the consistency of the running service network, we adopted the change management scheme proposed by Kramer and Magee \cite{R28}. In general, an element in a service network (including a process instance) can be in three management states: \textit{Active}, \textit{Passive}, and \textit{Quiescence}. The \textit{Passive} state of an element enables the system to complete the existing process instances, and to move the element to its \textit{Quiescence} state. If a runtime change to an element can adversely affect some existing process instances, then the change must be delayed until the element reaches its \textit{Quiescence} state. A newly added element is always in the \textit{Passive} state, and must be explicitly moved to the \textit{Active} state so that new process instances can use the element. An element can be removed from the system when it is in \textit{Quiescence} state. The events are generated at each management state change, allowing the organizational and operational managers to observe and make appropriate management decisions (i.e., triggering ECA rules).
\subsubsection {Specification and enactment of management policies}\hfill \\
To specify the organizational and operational management decisions, SDSN@RT provides a policy language by adopting the language of the Drools business rule engine. A policy consists of a set of ECA rules. 

\begin{itemize}
\item \textit{Conditions}. A policy rule needs to be able to react to service network events (e.g., a management state change), service network state changes (e.g., the unavailability of a service), the properties of VSNs and their instances, the time of the day or other context conditions, and the execution state of the management policies. Thus, a condition can be a logical expression of the predicates about the information in these sources. 
\item \textit{Actions}. The actions can include the above-mentioned organizational and operational management capabilities (e.g.,  add role, remove contract, and add routing rule). Additionally, they can be custom management functions, e.g., predicting response time violations. 
\item \textit{Execution}. The correct ordering of the rules as well as that of the actions within each rule are required to achieve a desired outcome. The rules are independent and are activated based on their conditions. When multiple rules are activated at the same time, the priorities of the rules can be used to resolve any conflicts. Within a rule, if-then-else conditional constructs can be used to order the actions. 
\end{itemize}

\indent The policy manager in the management platform uses a policy engine to enact the management policies. It supports addition, removal, and update of policies. It can parse given policies, process events, state changes, and other provided information, and activate the rules in the policies as their conditions are met.
\begin{figure}
\centering
\includegraphics [scale=.65,keepaspectratio]{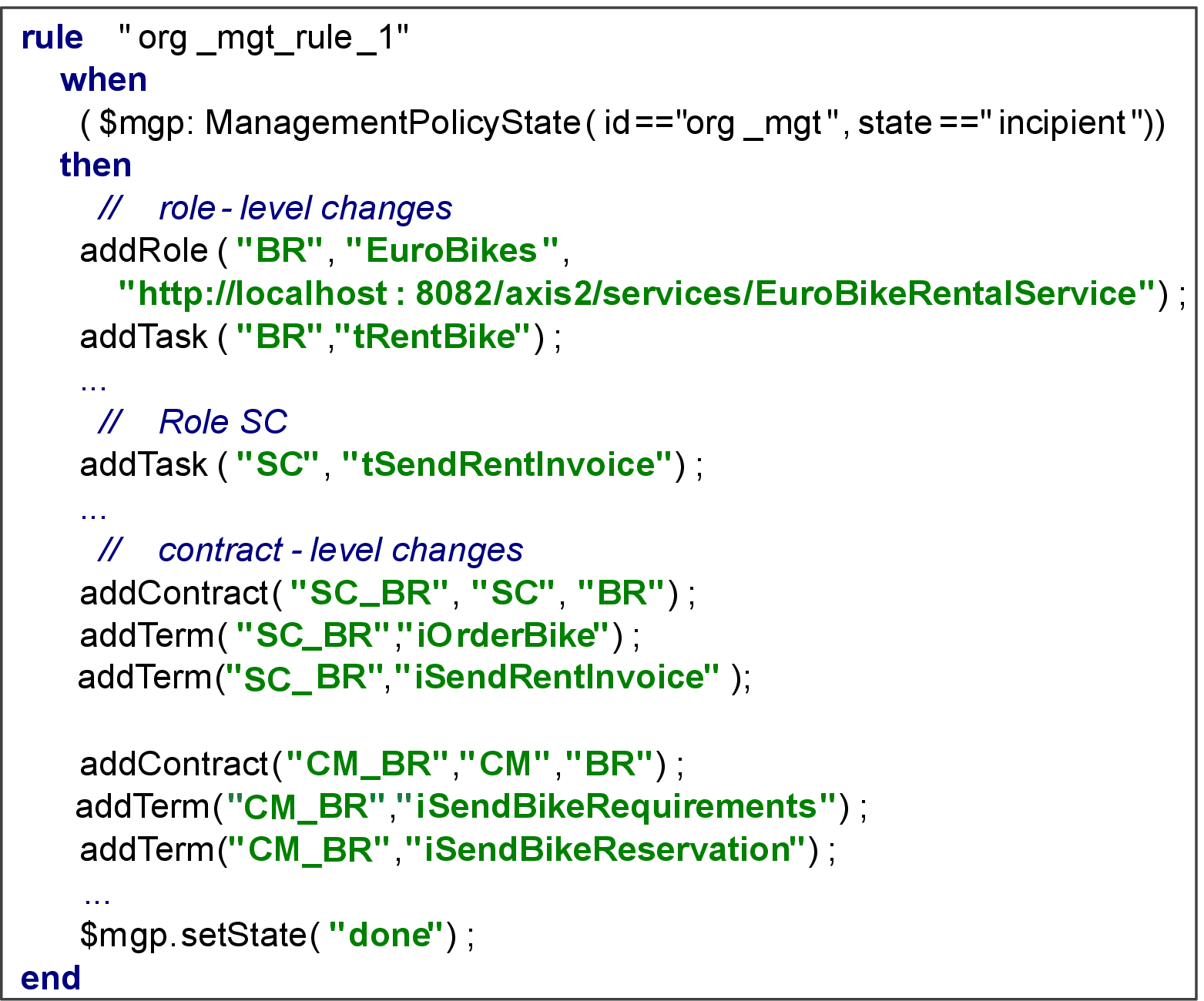}
\caption{A fragment of the organizational management policies for adding the bike rental feature}
\end{figure}
\begin{figure}
\centering
\includegraphics [scale=.65,keepaspectratio]{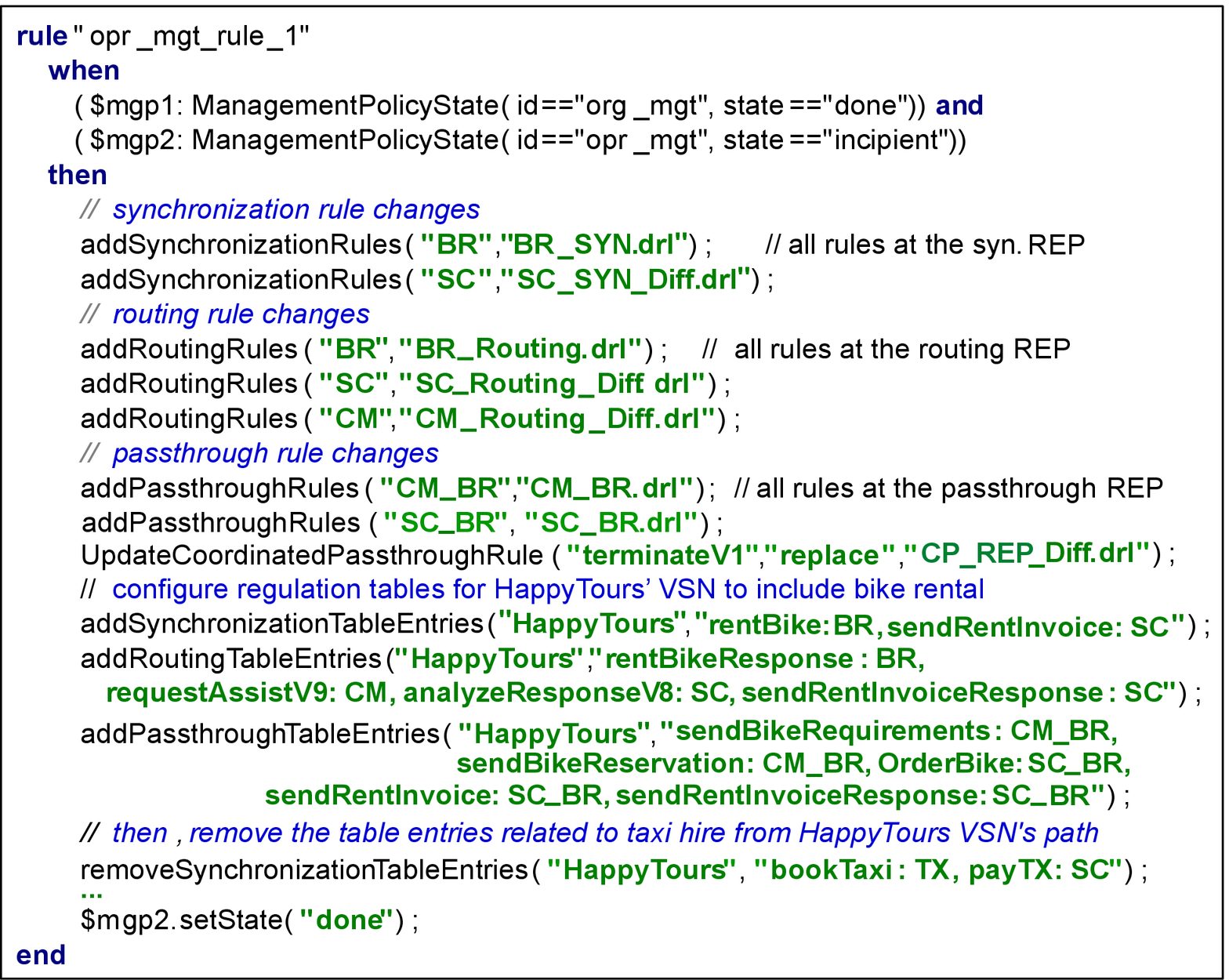}
\caption{A fragment of the operational management policies for adding the bike rental feature and configuring HappyTours's VSN to use it, replacing the taxi hire feature}
\end{figure}

\indent Figure 10 shows a fragment of the organizational management policy that alters the camping assistance service network by incorporating the implementation of the bike rental feature. The implementation is a collaboration between the bike rental service (EuroBikes), 24by7 support center, and the camper. To represent the bike rental service, a new role BR (a new node) is created in the service network. To represent the inter-connections among the participants in the bike rental collaboration, the contracts or links, such as the contract SC\_BR and CM\_BR, are added. The relevant tasks and the interaction terms are also included.

\indent Figure 11 shows a fragment of the corresponding operational management policy. The regulation rules that are required to support the managed interactions between the participants in the bike rental collaboration are added to the relevant REPs. The individual rules are defined as Drools rule files (.drl). As the tenant HappyTours needs to use the new bike rental feature, and to drop the taxi hire feature, the service network path of the HappyTours's VSN is modified by changing the relevant regulation tables (i.e., adding and removing table entries).
\subsection{Prototype implementation}
We built the prototype of the SDSN@RT middleware by adopting and further extending the ROAD/Serendip framework, which support adaptive service orchestrations \cite{R5, R12}. We used the Java programming language to develop the middleware. The prototype follows the system architecture shown in Figure 3. We have implemented the coordination infrastructure by incorporating the support for the regulation and management capabilities into the service composition engine of ROAD. We have also extended the design and management of the service orchestrations in ROAD to fully support the configuration design of a multi-tenant service network. Each node (role) in a service network is an Axis2 Web service, and uses the Axis2 Web service client API to send messages to, and receive messages from the external service. As a Web service, a node has a WSDL and an endpoint reference. The external clients/systems can send the messages to the node using its WSDL and endpoint reference.

We have implemented and added the management platform as a module to Apache Axis2 Web service engine (axis.apache.org/axis2). The Drools rule engine and its language were used to define and execute the regulation rules and management policies. The SDSN implementation (including all samples) is available at https://github.com/road-framework/SDSN. The size of the project has 407356  lines of code (as per Github GLOC on 3/11/2018), which include Java, XML, and Drools rule files. The base ROAD framework is also available at https://github.com/road-framework (ROADFactory and ROADWS sub-projects). The size of the base framework is 72173 in lines of code.
\section{Evaluation}
Previously in evaluating our approach, we have carried out evaluations on performance differentiation, capacity utilization, and comparison to the MIMT model \cite{R3}.  We have implemented and used three different throughput sharing schemes (selected from the existing works). Compared with the MIMT model, the SIMT model has been found to achieve a better utilization of the capacities or throughputs (e.g., the maximum number of possible repairs at a repair shop at a given moment) of the business services with little performance overhead (see Section 6.2 in \cite{R3}). In the remainder of this section, we focus on the evaluation of our middleware design.

\indent The evaluation of our middleware design consists of two case studies using our prototype, and a performance study of our prototype. The former shows that our middleware design can meet the requirements for a middleware that can host SIMT composite cloud applications, and facilitate the management of such applications. The latter evaluates the runtime overhead of the key capabilities of the middleware: deployment and hosting of SIMT applications, enactment of application variants, and management of SIMT applications. The case study resources are available at \textit{https://github.com/IndikaKuma/SDSN\_SPE\_2018}. 
\subsection{Case studies}
As a proof of concept, we have fully developed the camping assistance SIMT application presented in this paper, using our approach and middleware. The services are Apache Axis2 Web services. We have also developed a roadside assistance case study (RoSAS), which was presented in our earlier publications on SDSN design support \cite{R3,R31}. Tables 3 and 4 show the software metrics of both case studies. We have analyzed the case studies by observing the execution of VSNs and services, and by examining the logs generated by the middleware and the services (see the case study resources for the logs). All the following exercises (for both case studies) were successfully completed with our middleware support.
\begin{table}
\centering
\caption{Software metrics for the camping assistance (CampSAS) case study}
\begin{adjustbox}{max width=\textwidth}
\begin{tabular}{|l|l|l|}
\hline
\textbf{Artifact}       & \textbf{LoC (Lines of Code)}  & \textbf{Details}        \\ 
\hline 
Services  & 945 (Java) & 9 Services \\ 
\hline
SIMT Application & 1186 (XML), 639 (Drools) & 10 roles, 13 tasks, 17 contracts, 22 interaction terms, 78 regulation rules\\ 
\hline
Management Policies  & 197 (Drools)   & 15 rules \\ 
\hline
\end{tabular}
\end{adjustbox}
\end{table}
\begin{table}
\centering
\caption{Software metrics for the roadside assistance (RoSAS) case study}
\begin{adjustbox}{max width=\textwidth}
\begin{tabular}{|l|l|l|}
\hline
\textbf{Artifact}       & \textbf{LoC (Lines of Code)}  & \textbf{Details}        \\ 
\hline 
Services  & 1380 (Java) & 8 Services \\ 
\hline
SIMT Application & 1540 (XML), 950 (Drools) & 9 roles, 21 tasks, 15 contracts, 17 interaction terms, 115 regulation rules\\ 
\hline
Management Policies  & 280 (Drools)   & 18 rules \\ 
\hline
\end{tabular}
\end{adjustbox}
\end{table}

\begin{itemize}
\item \textit {Deployment of the service network.}
First, the middleware creates the topology of the CampSAS/RoSAS service network in terms of runtime objects of roles, tasks, contracts, and interaction terms. The REPs for the service network are also created, and subscribed to the event manager and the state manager. Second, the middleware populates the knowledgebases at REPs using the relevant regulation rules (Drools) that are defined as part of the regulation design of the CampSAS/RoSAS assistance service network.
\item \textit {Deployment of VSNs.} The middleware uses the VSN designs of the three tenants of CampSAS/RoSAS to populate the regulation tables at REPs with VSN-to-regulation-rule mappings. 
\item \textit {Enactment of VSNs.} Upon receiving a Web service request from a user of a tenant at the client role (the role CM in CampSAS service network), the enactment engine creates and executes an instance of the tenant's VSN. It can simultaneously enact the instances of the VSNs of multiple tenants on the same CampSAS/RoSAS service network, while ensuring the isolation among VSN instances. 
\item \textit {Runtime update of the multi-tenant service network.} We applied the organizational and operational management policies presented in Section 3.5 to the running CampSAS service network. For the RoSAS service network, the runtime evolution scenario was adding the taxi hire feature. The online case study resources include the complete management policies for both CampSAS and RoSAS. Via the logs and the response messages of VSN executions, we validated the controlled modification of the service network. The relevant logs in the case study resources are annotated to highlight this runtime behavior. We also compared the logs and VSN responses with those of the manually created same service network to validate the changes to the service network.  
\item \textit {Un-deployment of the VSNs and the service network.}
We first un-deployed the three VSNs one-by-one (via management policies), and then removed the service network (via the deployment interface).
\item \textit {Coverage of runtime changes.} To validate our support for different types of runtime changes to a multi-tenant service network, we created and updated the same service network by incrementally adding the realizations of 10 features, all at runtime. These features are \textit{Case Handling}, \textit{Reimbursement}, \textit{Towing}, \textit{Repairing}, \textit{Spare Parts}, \textit{Vehicle Repair Assessment}, \textit{Legal Assistance}, \textit{Accommodation}, \textit{Rental Vehicle}, and \textit{Taxi Hire}. These features are common to both the domains of camping assistance and roadside assistance, for example, repairing/towing a car/bike/caravan and reserving special accommodation for some campers. We also considered the rollback (removal) of each feature. In general, a feature is realized by a collaboration among a subset of the services. Adding a feature requires changes to both the topology and regulation designs of the service network. For example, the addition of the feature Case Handling requires 1 service and 1 client, 2 roles, 4 tasks, 1 contract, 1 interaction term, 1 VSN, and 12 regulation rules (for the four types of REPs). These scenarios together cover each type of service network changes at least one time. 

To realize a given feature, we first identify the differences between the current service network, and the target service network after the implementation of the feature, in terms of service network changes. Then, we create the organizational and operational management policies and the regulation rules, to realize the identified changes. Next, we apply the created policies at runtime on the current service network. We validated the implementation of the scenario following the same procedure used to validate the updates to the CampSAS/RoSAS service network (see above).  
\item \textit {Deployment and enactment of multiple service networks.} We deployed 10 different multi-tenant service networks on the same middleware instance, and simultaneously executed each of the VSNs in the service networks successfully. To create these service networks, we have incrementally used the above-mentioned 10 features. The k\textsuperscript{th} multi-tenant service network realizes 1\textsuperscript{st} to k\textsuperscript{th} features (1$\leq$k$\leq$10). Each service network has a single VSN whose enactment covers the whole service network (i.e., execute each task and rule).  
\end{itemize}
\subsection{Performance evaluation}
In these experiments, we quantify the absolute and relative performance overhead induced by our middleware design and its prototype. We consider the runtime overhead of deploying service networks and VSNs, enacting VSNs, and modifying service networks and VSNs. Our experiment setup consists of a workload generator (Apache AB), middleware (SDSN@RT), and Web services (Apache Axis2). Each system is deployed in a virtual machine with four 3400 MHz CPUs (Intel Broadwell), 8GB memory, and Ubuntu 16.04.4 LTS. Java version is Oracle JDK 1.8.0-161. To measure the memory usage, we used the top utility in Ubuntu.
\subsubsection{Deployment overhead} \hfill \\
We used the 10 different SIMT applications mentioned in Section 4.1. For each SIMT application, we deployed 50 copies of the service network of the SIMT application, and took the average deployment time and physical memory consumption. We used 100 copies of the VSN in the SIMT application to quantify the average deployment time and physical memory usage for VSNs.\\
\indent Figure 12 shows the deployment time and memory consumption for service networks with respect to their sizes in LoC (Lines of Code). Each of two measures increases approximately linearly. The size of a service network (approximately) proportionally corresponds to the number of runtime objects created during the deployment. Thus, as the size of the service network increases, its deployment time and memory usage also increases. \\
\indent In general, the deployment of a VSN simply adds a set of regulation table entries (the mappings between VSN identifiers and regulation rule identifiers). Thus, the deployment overheads for the VSNs are approximately similar. The deployment time of a VSN has the mean of 3.374 milliseconds and the standard deviation of 0.23 milliseconds. The memory usage has the mean of 12.76 KB and the standard deviation of 2.09 KB.\\
\begin{figure}
\centering
\includegraphics [scale=.65,keepaspectratio]{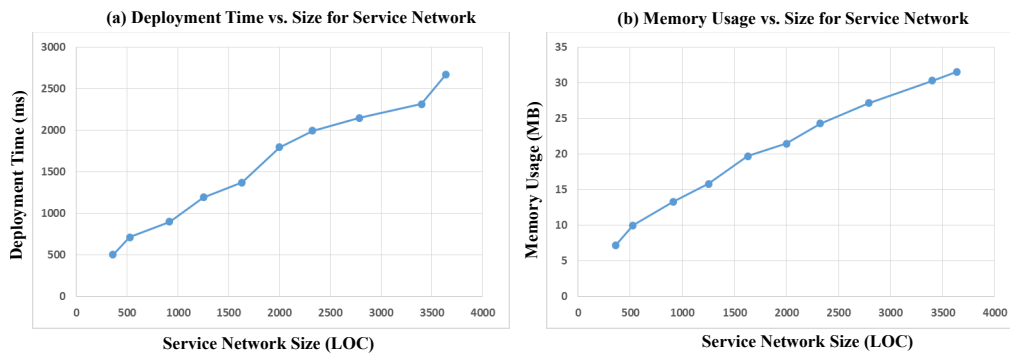}
\caption{Overhead of service network deployment (a) deployment time, (b) physical memory usage }
\end{figure}
\subsubsection{Enactment overhead} \hfill \\
To quantify the execution overhead for VSNs, we used the same 10 SIMT applications mentioned in Section 4.1. In these scenarios, the enactment of the VSN covers the entire service network for the SIMT application. As we only quantify the overhead included by our middleware, we exclude the service execution time from the enactment time of the VSN. We first issued 100 (with the concurrency level of 10) warm-up calls. The concurrency level is the number of multiple requests to perform at a time (see Apache AB documentation). Next, we issued 2000 calls (with the concurrency level of 100) and measured the average response time at the client side. Figure 13 shows the enactment time (across 100 concurrent requests) for VSNs against the size of the service network. The enactment overheads increase approximately linearly with the size of the service network path that the VSN covers. \\
\begin{figure}
\centering
\includegraphics [scale=.4,keepaspectratio]{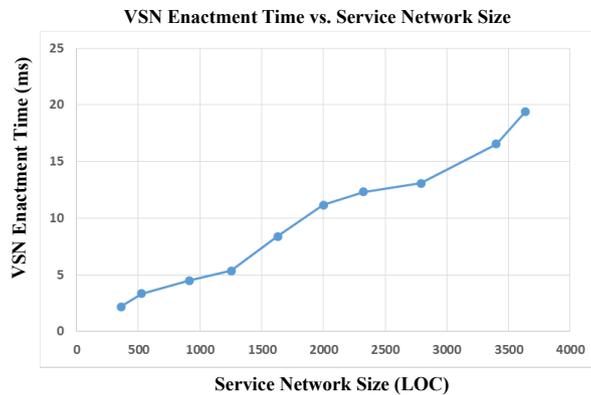}
\caption{Overhead of VSN enactment (across, 100 concurrent requests)}
\end{figure}
\subsubsection{Management overhead} \hfill \\
To quantify the runtime management overhead, we used the 10 change scenarios mentioned in Section 4.1, where each scenario adds a new feature. For each scenario, we have measured the run-time change enactment time (RCET), which is the time difference between the middleware receiving the management policies and its being ready for use after applying policies. As active VSN instances can delay the execution of the management policies, we enacted the policies when there are no active VSN instances. 

\indent Figures 14 and 15 show the RCET and size of management policies for the realization and rollback of changes. The size of the management policy also includes the regulation rules added or updated. For the 10 scenarios, the RCET for the realization of changes is within 730-939 milliseconds, and that for the rollback of the realizations is within 73-773 milliseconds. We believe that this is reasonable. In general, RCET depends on the number of atomic change operations (which, approximately corresponds to the size of the management policies) as well as the types of atomic change operations (addition, update, or removal).  The addition of an object (e.g., role or a regulation rule) is expensive compared to the removal of the corresponding object. The rollback of a change realization can involve the addition of removed objects, the removal of added objects, and the update of the changed objects. 
\begin{figure}
\centering
\includegraphics [scale=0.9,keepaspectratio]{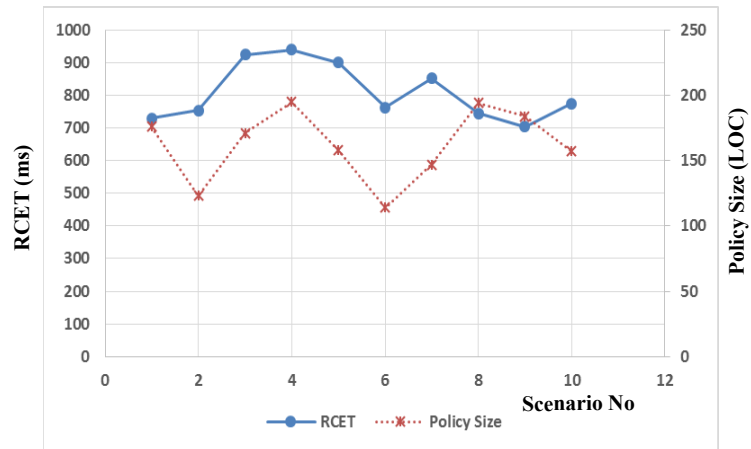}
\caption{Run-time change enactment time (RCET) and management policy size for change scenarios: feature addition or change realization}
\end{figure}
\begin{figure}
\centering
\includegraphics [scale=0.45,keepaspectratio]{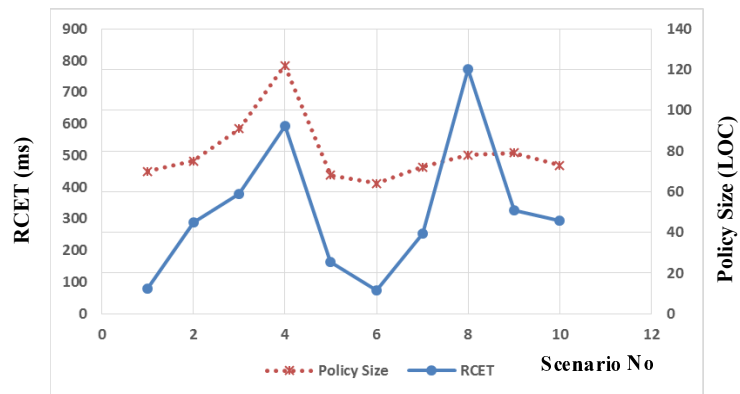}
\caption{Run-time change enactment time (RCET) and management policy size for change scenarios: feature removal or change rollback}
\end{figure}
\subsubsection{Relative enactment overhead} \hfill \\
In our earlier publications on our SDSN design support \cite{R3,R31}, we have compared the SIMT model in our SDSN approach and the SIMT model in BPEL (Business Process Execution Language) \cite{R32} with Apache ODE (ode.apache.org) in terms of runtime enactment overhead \footnote{This exercise was for comparing the middleware performance only. The SIMT model in BPEL does not have the dynamic sharing and adaptation capabilities of the SIMT model in our SDSN approach \cite{R3}}. We measured the end-to-end application enactment time (including service execution times) at the client side. We have used the roadside assistance case study in this experiment. Our SDSN middleware is shown to have similar enactment overhead to Apache ODE.
\section{Related Work}
The functional and performance differentiation in an SIMT composite application requires support for the runtime sharing and variation of services and the configuration design (architectural composition of services) and regulation design (regulation of conversational behaviors between services) of the application. A middleware for hosting SIMT applications need to be able to create and maintain the runtime models of the applications, simultaneously enact the application variants in a controlled manner, and enable the runtime management of the applications and their variants. This section compares our middleware with related works in middleware for component/service compositions, and multi-tenant applications.
\subsection{Middleware for composite applications}
Middleware have been proposed for two main types of composite applications: component-based applications and service-based applications. The former is a precursor to the latter. For a comparison of these two types of applications, the interested readers are referred to the relevant studies \cite{R33,R34,R35}. This research focuses on service-based applications. With service-orientation, the implementations of the services are hidden from service consumers (the services are black boxes). They have discoverable interfaces (e.g., WSDL), and are consumed by sending messages through Web technologies. The composition of services are done through coordinating interactions (message exchanges) between services. The main coordination model types are orchestration and choreography. The orchestration model requires a special middleware, and this paper has presented a service orchestration middleware for service-based single-instance multi-tenant cloud applications.

Most middleware systems for component-based applications \cite{R4, R5, R6, R36, R37} can deploy a specification of a component assembly. The deployment generally creates runtime models. These middleware systems also provide management capabilities for these runtime models such as intercepting and monitoring message exchanges between components, injecting extensions/plug-ins during invocation at predefined hooks (extension points), and modifying runtime models. The component models supported by these middleware systems include a subset of the common software architectural abstractions such as components, connectors, composites, and (provided and required) interfaces.

\indent Similarly, most existing middleware systems for composite service applications \cite{R38, R39, R40} can enact a specification of the composite service by creating runtime models for representing the composite service. The runtime management capabilities such as discovering services, monitoring message exchanges between services, replacing services, modifying the definition and instances of the composite applications, and providing quality of service (QoS) for services by managing computing resources used by the middleware. Common service composite models supported by these middleware systems include process-centric models (e.g., BPEL or BPMN (Business Process Model and Notation)\cite{R41}), component-based models \cite{R4}, and organization-based models \cite{R5}. The process-centric models consider how business activities or interactions are organized and ordered to achieve business goals. Component-based models define composite applications as assemblies of components that provide and request services. Organization-based models represent service compositions as goal-oriented organizations using roles, players realizing roles, interaction and normative relationships between roles, and controller(s) managing the organizations.

\indent The service network model can also be considered as a service composition model. It can provide a natural abstraction to design, enact, regulate, and manage webs of real-world business service networks \cite{R26,R27}. Most existing works on service networks consider the modeling and analysis of service networks from specific aspects \cite{R27} such as business value flow \cite{R42}, and service relationships \cite{R26}. Their realizations have relied on process-centric models, which fail to represent service networks naturally. That is, the domain concepts and their representations are mismatched, and the domain concepts (e.g., services, service capabilities, service relationships, service interactions, interaction routing and regulation, service network paths, and virtualization) are not directly represented or managed in the realization, limiting their utility. To the best of our knowledge, SDSN@RT is the first middleware to support the deployment, enactment, and management of service-based applications designed following the (extended) service network model. Note that, recently, the concept of service mesh \cite{R43,R44} was used to describe the inter-service communication infrastructure for microservices. However, compared with the service network model, the service mesh model does not have the abstractions required to represent the above-mentioned domain concepts of multi-tenant business service networks.  
\subsection{Middleware for multi-tenant applications}
There exist middleware systems for multi-tenant service-based applications as well as multi-tenant component-based applications. They can be further categorized based on two multi-tenant models: MIMT and SIMT. Their capabilities and approaches to support functional and performance differentiation for tenants also differ. A few of these works support runtime changes to multi-tenant applications. 

\indent Some process engines \cite{R9} and enterprise service buses (ESB)\cite{R45} implement the MIMT model for composite service applications. The service provider hosts the middleware, and tenants create and manage their own services and processes. The tenant-specific meta-data are used to customize software elements (services or ESB components), and each created variant is separately deployed and managed (i.e., the MIMT model). Tenants can define their applications (processes or message mediation flows) that use their own customized software elements. To achieve the performance differentiation among tenants, the regulation functions such as global admission controllers or schedulers are used to fairly allocate middleware-level computing resources among tenants \cite{R17}.

We identify three approaches to enacting the variant instances of an SIMT composite application for different tenants and their users: generative approach, dynamic-binding approach, and virtualization (slicing) approach.

The generative approach \cite{R16, R46, R47} derives application variants by customizing the composite application for each user request, and deploys and enacts the derived variant.  This approach can utilize the existing works on the customization support for composite applications. In our earlier works, we also supported customizing service compositions \cite{R48} and multi-tenant applications \cite{R49}. The composite service application is represented as a customizable application, which includes all allowed variations. The model-driven engineering techniques are used to resolve variability, and to generate executable application variants. An alternative approach is to select a subset of application fragments (e.g., process fragments \cite{R50} or Java code fragments \cite{R51}), and merge them to form the application variant. The generative approach can potentially incur a considerable runtime overhead (variant generation time + deployment and compilation time + enactment time) as the variability and size of the composite cloud application increase. Changing the design of the application variant while it is being enacted requires the capability to modify the running application instance or regeneration of a new application variant and migrating the running instance to the new variant.

In the dynamic-binding approach \cite {R8,R10,R15}, specific elements are selected, attached, and executed at the extension points of the application, for example, dynamically selecting/binding a service, executing ECA rules, selecting a control branch in a business process, and injecting a component via dependency injection. At the design time, a software engineer needs to annotate the extension points in the application specification, and codify the mappings between tenant requirements/features, extension points, and choices at each extension points. The middleware needs to deploy such application specifications, and enact application variants at runtime by resolving these mappings (runtime variability resolution). A tenant selects a set of extension points and the choices for them. Using tenant information, the composite application identifies the extension points for the tenant, and binds/selects the relevant choices at them during the execution of the application.

There are several limitations with the existing works that use the dynamic-binding approach. First, a tenant requirement or feature is generally not mapped to a single component or service but to a collaboration among a set of components or services \cite {R3, R13, R52}. Thus, activating/deactivating a feature requires activating/deactivating a collaboration, which requires the ability to dynamically change the architectural topology of the application variant and the conversational behaviors within it. This capability is not sufficiently supported by the existing works. Second, the performance differentiation requires the regulation of interactions (message or object exchanges) between services/components to dictate progression of the instances of the application variants. The application should be able to execute tenant-specific regulation policies (with appropriate sharing of policies between tenants) to enforce the tenant-specific regulations on each service interaction. These capabilities are also not sufficiently available in the related studies. Third, most of these approaches annotate the variation points and/or their options within the composite application at design time, which can increase the design complexity and reduce the maintainability of the application \cite{R13}. For example, it is difficult to represent modular abstractions such as collaborations, and it involves the tangling of normal processing logic with variant differentiation logic.

In the virtualization approach \cite{R12}, the complete virtual variants of the shared composite application are formed per tenant and/or per user request. The middleware maintains the runtime model of the application (models@runtime). The enactment of an application instance executes a subset of the elements of the shared runtime model of the application (no recompilation or changes to the runtime model). In one of our earlier works \cite{R12}, the virtualization approach was applied to realize the SIMT model for the composite cloud applications designed following the organization-based service composition model. The support for runtime sharing with variations was limited to the configuration design of the application. Moreover, an assumption was made that changing service providers (role-players) can be done by changing role-player/service bindings, which is not always valid for business services.

To the best of our knowledge, we are the first to support the SIMT model for composite service applications with the virtualization approach and the service network model. We extend the service network model with virtualization and regulatory control of the conversational behaviors in the service network. SDSN@RT creates and maintains the runtime model of a multi-tenant service network. This runtime model acts as an overlay network over the external services. The regulation enforcement points (REPs) in the overlay network makes the message routing over it explicit, visible, and controllable at the appropriate granularity. With managed dynamic routing, virtual service networks can be dynamically formed per tenant and user request. The organizational and operational managers of the service network can change its topology and the routing and regulation logic at REPs at runtime. This virtualization of a composite application is analogous to the virtualization of a computer network. Virtual networks share the same physical computer network, and each virtual network has a sub topology of the physical network, and the network packets are routed and regulated in the sub topology according to the policies defined for the virtual network. Some approaches to SDN (software-defined networking) based network virtualization apply the concept of slicing a physical network via controlled and programmable routing of network data packets \cite{R53}. 

Intercepting, regulating, and routing message/object/data exchanges between applications have been used by middleware systems for component-based applications \cite{R6,R54} and service-based applications \cite{R4,R55}. In general, these works focus on dynamically selecting a specific service/component, intercepting method calls to add crosscutting functionalities, or regulating messages to a specific service/component. However, in SDSN@RT, dynamic routing is used to form complete managed virtual application variants in an SIMT composite service application. It creates the virtual slices of services and the configuration and regulation designs of the composite service application to support functional and performance requirements of different tenants. Moreover, in the traditional reflective middleware for distributed object systems, the interceptors are generally registered at the middleware-level to intercept method calls of any application. In SDSN@RT, each composite application has its own intercepting and routing capabilities, which can be specified in the application design by a software engineer. The sharing and variations of these capabilities among multiple tenants/application variants are defined during the application design, and enforced during execution by the middleware.

Moreover, as mentioned above, in general, a collaboration among a set of services realizes a high-level tenant requirement or a feature.  Thus, to form an application variant for a tenant, a set of collaborations need to be composed/wired, and the interactions in the collaborations need to be regulated (locally and globally). Furthermore, the business services are heterogeneous. Even the services in the same category (see our motivating example) can have different capabilities, capacities, and relationships with other services. Thus, replacing a service cannot be done simply by changing the service binding. The message routing in the application needs to be altered. If an existing service needs to be replaced with a new service, the runtime model of the application needs to be modified structurally. SDSN@RT uses the models@runtime, service network model, and managed dynamic routing (via the regulation enforcement points) in the service network to support the creation of virtual service networks (application variants) by composing regulated and managed collaborations (between business services) dynamically. The domain specific language of SDSN@RT supports the design of a SIMT application following a compositional approach, where the configuration and regulation designs of the service network are decomposed into modular units by using the collaboration over services as the abstraction (see \cite{R3} for more information). Through the improved modularity, the compositional approach can potentially alleviate the drawbacks of the annotative approach \cite{R13}. 

\indent To support the runtime evolution of a multi-tenant composite cloud application, the middleware needs to support the potential classes of runtime changes to each key abstraction in the application. Moreover, the runtime models (models@runtime) of the application needs to be maintained to enable changes to the application without redeploying, recompiling, or halting it. The existing works \cite{R56, R57, R58} are limited to the changes in configuration designs, and do not support the fine-grained changes to the runtime models of regulation designs. They support changes at the level of the choices for the variation points such as the addition and removal of choices, versioning of services or process definitions, the addition and removal of alternative services of the same type, and the modification of service dependencies. 

\begin{table}[]
\centering
\caption{A Summary of the Comparative Analysis of the Related Works (Including Our SDSN@RT Approach)}
\begin{adjustbox}{max width=\textwidth}
\begin{tabular}{llllllllll}
\hline
\multicolumn{1}{|l|}{Criteria \textbackslash Approach}                                                 & \multicolumn{1}{l|}{}                                                                                       & \multicolumn{1}{l|}{[8]}      & \multicolumn{1}{l|}{[12]}       & \multicolumn{1}{l|}{[16]}                                                  & \multicolumn{1}{l|}{[10]}                                                  & \multicolumn{1}{l|}{[15]}                                                  & \multicolumn{1}{l|}{[43]}     & \multicolumn{1}{l|}{[9]}                                                   & \multicolumn{1}{l|}{SDSN@RT}                                                                                                                                                                                             \\ \hline
\multicolumn{1}{|l|}{\begin{tabular}[c]{@{}l@{}}Service/Component\\ Orientation\end{tabular}}          & \multicolumn{1}{l|}{}                                                                                       & \multicolumn{1}{l|}{CO}     & \multicolumn{1}{l|}{SO}       & \multicolumn{1}{l|}{SO}                                                  & \multicolumn{1}{l|}{SO}                                                  & \multicolumn{1}{l|}{SO}                                                  & \multicolumn{1}{l|}{SO}     & \multicolumn{1}{l|}{SO}                                                  & \multicolumn{1}{l|}{SO}                                                                                                                                                                                                  \\ \hline
\multicolumn{1}{|l|}{\begin{tabular}[c]{@{}l@{}}Service \\ Composition \\ Model\end{tabular}}          & \multicolumn{1}{l|}{}                                                                                       & \multicolumn{1}{l|}{-}      & \multicolumn{1}{l|}{OR}       & \multicolumn{1}{l|}{\begin{tabular}[c]{@{}l@{}}PC\end{tabular}} & \multicolumn{1}{l|}{\begin{tabular}[c]{@{}l@{}}PC\end{tabular}} & \multicolumn{1}{l|}{\begin{tabular}[c]{@{}l@{}}PC\end{tabular}} & \multicolumn{1}{l|}{SM}     & \multicolumn{1}{l|}{\begin{tabular}[c]{@{}l@{}}PC\end{tabular}} & \multicolumn{1}{l|}{\begin{tabular}[c]{@{}l@{}}Service network model \\ extended with \\ collaborations, processes, \\ virtualization, regulation of \\ conversational behaviors,\\ and management/control\end{tabular}} \\ \hline
\multicolumn{1}{|l|}{\begin{tabular}[c]{@{}l@{}}Multi-Tenancy \\ Model\end{tabular}}                   & \multicolumn{1}{l|}{}                                                                                       & \multicolumn{1}{l|}{SIMT}   & \multicolumn{1}{l|}{SIMT}     & \multicolumn{1}{l|}{MIMT}                                                & \multicolumn{1}{l|}{SIMT}                                                & \multicolumn{1}{l|}{SIMT}                                                & \multicolumn{1}{l|}{MIMT}   & \multicolumn{1}{l|}{MIMT}                                                & \multicolumn{1}{l|}{SIMT}                                                                                                                                                                                                \\ \hline
\multicolumn{1}{|l|}{\begin{tabular}[c]{@{}l@{}}Runtime Variation \\ Mechanism\end{tabular}}           & \multicolumn{1}{l|}{}                                                                                       & \multicolumn{1}{l|}{DB}     & \multicolumn{1}{l|}{VR}       & \multicolumn{1}{l|}{CU}                                                  & \multicolumn{1}{l|}{DB}                                                  & \multicolumn{1}{l|}{DB}                                                  & \multicolumn{1}{l|}{DB}     & \multicolumn{1}{l|}{-}                                                   & \multicolumn{1}{l|}{\begin{tabular}[c]{@{}l@{}}Virtualization (VR)\\ via dynamic routing\end{tabular}}                                                                                                                        \\ \hline
\multicolumn{1}{|l|}{\multirow{2}{*}{Services}}                                                        & \multicolumn{1}{l|}{\begin{tabular}[c]{@{}l@{}}Sharing with \\ Variations\end{tabular}}                     & \multicolumn{1}{l|}{-}      & \multicolumn{1}{l|}{$\sim$}   & \multicolumn{1}{l|}{$\sim$}                                              & \multicolumn{1}{l|}{+}                                                   & \multicolumn{1}{l|}{-}                                                   & \multicolumn{1}{l|}{$\sim$} & \multicolumn{1}{l|}{$\sim$}                                              & \multicolumn{1}{l|}{+}                                                                                                                                                                                                   \\ \cline{2-10} 
\multicolumn{1}{|l|}{}                                                                                 & \multicolumn{1}{l|}{Business Services}                                                                      & \multicolumn{1}{l|}{-}      & \multicolumn{1}{l|}{$\sim$}   & \multicolumn{1}{l|}{-}                                                   & \multicolumn{1}{l|}{$\sim$}                                              & \multicolumn{1}{l|}{$\sim$}                                              & \multicolumn{1}{l|}{$\sim$} & \multicolumn{1}{l|}{$\sim$}                                              & \multicolumn{1}{l|}{+}                                                                                                                                                                                                   \\ \hline
\multicolumn{1}{|l|}{\multirow{2}{*}{\begin{tabular}[c]{@{}l@{}}Configuration \\ Design\end{tabular}}} & \multicolumn{1}{l|}{Models@runtime}                                                                         & \multicolumn{1}{l|}{$\sim$} & \multicolumn{1}{l|}{+}        & \multicolumn{1}{l|}{-}                                                   & \multicolumn{1}{l|}{$\sim$}                                              & \multicolumn{1}{l|}{$\sim$}                                              & \multicolumn{1}{l|}{$\sim$} & \multicolumn{1}{l|}{-}                                                   & \multicolumn{1}{l|}{+}                                                                                                                                                                                                   \\ \cline{2-10} 
\multicolumn{1}{|l|}{}                                                                                 & \multicolumn{1}{l|}{\begin{tabular}[c]{@{}l@{}}Sharing with \\ Variations\end{tabular}}                     & \multicolumn{1}{l|}{$\sim$} & \multicolumn{1}{l|}{+}        & \multicolumn{1}{l|}{-}                                                   & \multicolumn{1}{l|}{$\sim$}                                              & \multicolumn{1}{l|}{-}                                                   & \multicolumn{1}{l|}{-}      & \multicolumn{1}{l|}{-}                                                   & \multicolumn{1}{l|}{+}                                                                                                                                                                                                   \\ \hline
\multicolumn{1}{|l|}{\multirow{4}{*}{Regulation Design}}                                               & \multicolumn{1}{l|}{Models@runtime}                                                                         & \multicolumn{1}{l|}{-}      & \multicolumn{1}{l|}{-}        & \multicolumn{1}{l|}{-}                                                   & \multicolumn{1}{l|}{-}                                                   & \multicolumn{1}{l|}{$\sim$}                                              & \multicolumn{1}{l|}{$\sim$} & \multicolumn{1}{l|}{-}                                                   & \multicolumn{1}{l|}{+}                                                                                                                                                                                                   \\ \cline{2-10} 
\multicolumn{1}{|l|}{}                                                                                 & \multicolumn{1}{l|}{Granularity}                                                                            & \multicolumn{1}{l|}{-}      & \multicolumn{1}{l|}{-}        & \multicolumn{1}{l|}{-}                                                   & \multicolumn{1}{l|}{-}                                                   & \multicolumn{1}{l|}{FG}                                                  & \multicolumn{1}{l|}{FG}     & \multicolumn{1}{l|}{-}                                                   & \multicolumn{1}{l|}{FG}                                                                                                                                                                                                  \\ \cline{2-10} 
\multicolumn{1}{|l|}{}                                                                                 & \multicolumn{1}{l|}{\begin{tabular}[c]{@{}l@{}}Sharing with\\  Variations\end{tabular}}                     & \multicolumn{1}{l|}{-}      & \multicolumn{1}{l|}{-}        & \multicolumn{1}{l|}{-}                                                   & \multicolumn{1}{l|}{-}                                                   & \multicolumn{1}{l|}{-}                                                   & \multicolumn{1}{l|}{-}      & \multicolumn{1}{l|}{-}                                                   & \multicolumn{1}{l|}{+}                                                                                                                                                                                                   \\ \cline{2-10} 
\multicolumn{1}{|l|}{}                                                                                 & \multicolumn{1}{l|}{\begin{tabular}[c]{@{}l@{}}Integration with \\ Configuration \\ Structure\end{tabular}} & \multicolumn{1}{l|}{-}      & \multicolumn{1}{l|}{-}        & \multicolumn{1}{l|}{-}                                                   & \multicolumn{1}{l|}{-}                                                   & \multicolumn{1}{l|}{$\sim$}                                              & \multicolumn{1}{l|}{$\sim$} & \multicolumn{1}{l|}{-}                                                   & \multicolumn{1}{l|}{+}                                                                                                                                                                                                   \\ \hline
\multicolumn{2}{|l|}{Runtime Change Support (SIMT model)}                                                                                                                                                            & \multicolumn{1}{l|}{-}      & \multicolumn{1}{l|}{$\sim$}   & \multicolumn{1}{l|}{-}                                                   & \multicolumn{1}{l|}{-}                                                   & \multicolumn{1}{l|}{-}                                                   & \multicolumn{1}{l|}{-}      & \multicolumn{1}{l|}{-}                                                   & \multicolumn{1}{l|}{+}                                                                                                                                                                                                   \\ \hline
\multicolumn{2}{|l|}{Middleware DSL  (SIMT Model)}                                                                                                                                                                   & \multicolumn{1}{l|}{-}      & \multicolumn{1}{l|}{$\sim$CV} & \multicolumn{1}{l|}{-}                                                   & \multicolumn{1}{l|}{+ AV}                                                & \multicolumn{1}{l|}{$\sim$AV}                                            & \multicolumn{1}{l|}{-}      & \multicolumn{1}{l|}{-}                                                   & \multicolumn{1}{l|}{+ CV}                                                                                                                                                                                                \\ \hline
\multicolumn{10}{l}{}                                                                                                                                                                                                                                                                                                                                                                                                                                                                                                                                                                                                                                                                                                                                                                                                                                   \\
\multicolumn{10}{l}{\begin{tabular}[c]{@{}l@{}}+ Supported,   - Not Supported,   $\sim$Partially Supported,   DSL - Domain Specific Language\\ DB - Dynamic-Binding,  VR - Virtualization/Slicing, CU - Customization,  PC - Process-Centric, \\ CB - Component-Based,  SM - Service Mesh, OR - Organization-Based,  GB - Global Regulation\\ FG - Fine-Grained Regulation,  AV - Annotative Variability,  CV - Compositional Variability\end{tabular}}                                                                                                                                                                                                                                                                                                                                                                                                
\end{tabular}
\end{adjustbox}
\end{table}
Table 5 summarizes the comparison of SDSN@RT with the related works. SDSN@RT realizes the SIMT model for business service compositions. It supports the deployment, enactment, and runtime change management of SIMT composite applications that are modelled using the extended service network model. An SIMT application is represented as a multi-tenant service network, where a set of virtual service networks (per each tenant) coexists on the same service network at runtime. The regulation enforcement points (REPs) at the service network make the message passing over the service network observable and controllable at the appropriate granularity. At runtime, the application variants with different configuration and regulation designs are dynamically formed via routing, slicing the runtime model of the service network. The network view or controlled message passing between entities can enhance the capability to control conversational behaviors (interactions and message exchanges among participants) and dynamically form application variants in composite applications, thus supporting such management goals as virtualization and dynamic policy enforcement (as in flexible network-based models \cite{R23, R24} and dynamic routing in computer networks). SDSN@RT covers the potential classes of runtime changes to each of the key architectural abstractions in a multi-tenant service network. The architectural abstractions are represented explicitly using the models@runtime concept to enable the runtime changes to them. A software engineer can specify the desired runtime changes as policies consist of ECA rules, and enact the created policies at runtime without halting the running service network (or the SIMT application). 
\section{Conclusion and Future Work}
\noindent SDSN@RT is a middleware system that can deploy, enact, and manage single-instance multi-tenant (SIMT) composite SaaS applications. SDSN@RT represents an SIMT application as a multi-tenant service network, where a set of virtual service networks (VSNs) simultaneously coexist on the same physical service network at runtime. Each VSN realizes the functional and performance requirements of a particular tenant while sharing the underlying service network with other VSNs. The service network architecturally connects a set of services, and enables disciplined and fine-grained regulation of the interactions between services. VSNs may overlap as well as differ in terms of services and their capabilities, service network topologies, and regulation policies being applied to the interaction messages passing over the topologies. SDSN@RT can create and maintain the runtime models of the service network, and enact VSNs for multiple tenants over the shared service network simultaneously and in a controlled manner. It also facilitates the runtime changes to the service network and its VSNs. We have evaluated SDSN@RT and its support for enacting and managing SIMT composite cloud applications with a prototype, a case study, and a performance study. The experiment results have shown that SDSN@RT can deploy, and execute SIMT applications, and support the runtime modifications to the deployed applications, all with little performance overhead. 

\indent We have been exploring further such issues as regulation functions, and strategies for enacting VSNs with the optimal service sharing while meeting tenant expectations, VSN instance migration, runtime evolution of the multi-tenant service networks for performance-related changes (e.g., performance violation and abnormal termination of VSN instances), and high-level management of multi-tenant service networks (e.g., high-level abstractions and policies). Another future work is to improve SDSN@RT to support the deployment, enactment, and management of multi-tenant service networks in data centers with greater scalability and efficiency. We also plan to explore further aspects of SDSN@RT as our approach evolves over time, especially through the research insights concerning the dynamic management of communication networks from the SDN (software-defined networking)\cite{R23} and conventional computer network communities. We also plan to apply SDSN (or SDN concepts) to compose and manage microservices in a way that a logically centralized control is realized over choreography-style decentralized service interactions.

\section*{acknowledgements}
This research was partly supported by the Smart Services Cooperative Research Center (CRC) through the Australian Government's CRC Program, and by Swinburne University of Technology, and by the European Commission grant no. 825480 (H2020), SODALITE.


\begin{thebibliography}{99}
\bibitem{R1}Guo~CJ, Sun~W, Huang~Y, Wang~ZH, Gao~B. A framework for native multi-tenancy application development and management. \emph{Proceedings of 4th IEEE International Conference on Enterprise Computing, E-Commerce, and E-Services}, 2007; 551-55.
\bibitem{R2}Chong~F, Carraro~G. Architecture strategies for catching the long tail. \emph{MSDN Library}, Microsoft Corporation, 2006.
\bibitem{R3}Kumara~I, Han~J, Colman~A, Kapuruge~M. Software-Defined Service Networking: Performance Differentiation in Shared Multi-Tenant Cloud Applications. \emph{IEEE Transactions on Services Computing} 2017; \textbf{10}(1):9-22.
\bibitem{R4}Seinturier~L, Merle~P, Rouvoy~R, Romero~D, Schiavoni~V, Stefani~JB. A component-based middleware platform for reconfigurable service-oriented architectures. \emph{Softw. Pract. Exp.} 2011; \textbf{42}(5):559-583.
\bibitem{R5}Colman~A, Han~J. Using role-based coordination to achieve software adaptability. \emph{Science of Computer Programming} 2007; \textbf{64}(2):223-245.
\bibitem{R6}Truyen~E, Vanhaute~B, J\o rgensen~BN, Joosen~W, Verbaeten~P. Dynamic and selective combination of extensions in component-based applications. \emph{Proceedings of ICSE '01: International Conference on Software Engineering}, 2001; 233-242.
\bibitem {R7}Truyen~E, Cardozo~N, Walraven~S, Vallejos~J, Gunther~S, Joosen~W. Context-oriented programming for customizable SaaS applications, \textit{Proceedings of the 27th annual ACM symposium on applied computing}, 2012; 126-137.
\bibitem{R8}Walraven~S, Truyen~E, Joosen~W. A middleware layer for flexible and cost-efficient multi-tenant applications. \emph{Proceedings of the 12th International Middleware Conference}, 2011; 370-389.
\bibitem{R9}Pathirage~M, Perera~S, Kumara~I, Weerasiri~D, Weerawarana~S. A scalable multi-tenant architecture for business process executions. \emph{International Journal of Web Services Research} 2012; \textbf{9}(2):21-41.
\bibitem{R10}Gey~F, Walraven~S, Van~Landuyt~D, Joosen~W. Building a customizable business-process-as-a-service application with current state-of-practice. \emph{Proceedings of the 12th International Conference on Software Composition}, 2013; 113-127.
\bibitem{R11}Mietzner~R, Leymann~F, Papazoglou~M. Defining composite configurable SaaS application packages using SCA, variability descriptors and multi-tenancy patterns. \emph{Proceedings of Third International Conference on Internet and Web Applications and Services}, 2008; 156-161.
\bibitem{R12}Kapuruge~M, Han~J, Colman~A. \emph{Service Orchestration as Organization: Building Multi-Tenant Service Applications in the Cloud}. Morgan Kaufmann: San Mateo, 2014.
\bibitem{R13}Kastner~C, Apel~S, Kuhlemann~M. Granularity in software product lines. \emph{Proceedings of the 30th international conference on Software engineering}, 2008; 311-320.
\bibitem{R14}Reichert~M, Hallerbach~A, Bauer~T. Lifecycle management of business process variants. \emph{In Handbook on Business Process Management}, 2015; 251-278.
\bibitem{R15}MingXue~W, Bandara~K, Pahl~C. Process as a service distributed multi-tenant policy-based process runtime governance. \emph{Proceedings of 2010 IEEE International Conference on Services computing}, 2010; 578-585.
\bibitem{R16}Geebelen~K, Walraven~S, Truyen~E, Michiels~S, Moens~H, De Turck~F, Dhoedt~B, Joosen~W. An open middleware for proactive QoS aware service composition in a multi-tenant SaaS environment. \emph{Proceedings of the International Conference on Internet Computing}, 2012; 111-117.
\bibitem{R17}Patikirikorala~T, Kumara~I, Colman~A, Han~J, Wang~L, Weerasiri~D, Ranasinghe~W. Dynamic performance management in multi-tenanted business process servers using nonlinear control. \emph{Proceedings of 10th International Conference on Service-Oriented Computing}, 2012; 206-221.
\bibitem{R18}Hailue~L, Kai~S, Shuan~Z, Yanbo~H. Feedback-control based performance regulation for multi-tenant applications. \emph{Proceedings of 15th International Conference on Parallel and Distributed Systems}, 2009; 134-141.
\bibitem{R19}Krebs~K, Momm~C, Kounev~S. Metrics and techniques for quantifying performance isolation in cloud environments. \emph{Science of Computer Programming} 2014; \textbf{90}:116-134.
\bibitem{R20}Li~X, Liu ~TC, Li~Y, Chen~Y. SPIN: Service performance isolation infrastructure in multi-tenancy environment. \emph{Proceedings of the 6th International Conference on Service-Oriented Computing}, 2008; 649-663.
\bibitem{R21}Krebs~K, Loesch~M, Kounev~S. Platform-as-a-service architecture for performance isolated multi-tenant applications. \emph{Proceedings of IEEE 7th International Conference on Cloud Computing (CLOUD)}, 2014; 914-921.
\bibitem{R22}Blair~G, Bencomo~N, France~R. Models@run.time. \emph{Computer} 2009; \textbf{42}(10): 22-27.
\bibitem{R23}Kreutz~D, Ramos~FM, Verissimo~PE, Rothenber~CE, Azodolmolky~S, Uhlig~S. Software-defined networking: A comprehensive survey. \emph{Proceedings of IEEE} 2015; \textbf{103}(1):14-76.
\bibitem{R24}Kagal~L, Finin~T. Modeling conversation policies using permissions and obligations. \emph{Autonomous Agents and Multi-Agent Systems} 2007; \textbf{14}(2): 187-206.
\bibitem{R25}Kramer~J, Magee~J. Dynamic configuration for distributed systems. \emph{IEEE Transactions on Software Engineering} 1985; \textbf{11}(4):424-436.
\bibitem{R26}Danylevych~O, Karastoyanova~D, Leymann~F. Service Networks Modeling: An SOA \& BPM Standpoint. \emph{Journal of Universal Computer Science} 2010; \textbf{16}(13):1668-1693.
\bibitem{R27}Razo-Zapata~IS, Leenheer~P, Gordijn~J, Akkermans~H. Service Network Approaches. \emph{Handbook of Service Description: USDL and its Methods}, 2012; 45-74.
\bibitem{R28}Kramer~J, Magee~J. The evolving philosophers problem: dynamic change management. \emph{IEEE Transactions on Software Engineering} 1990; \textbf{16}(11):1293-1306.
\bibitem{R29}Forgy~CL. Rete: A fast algorithm for the many pattern/many object pattern match problem. \emph{Readings in Artificial Intelligence and Databases}, 1988; 547-559.
\bibitem{R30} Scheer~A, Thomas~O, Adam~O. Process Modeling using Event-Driven Process Chains. \emph{Process-aware information systems: bridging people and software through process technology}, 2005; 119-145. 
\bibitem{R31}Kumara~I, Han~J, Colman~A, Kapuruge~M, Software-Defined Service Networking: Runtime Sharing with Performance Differentiation in Multi-Tenant SaaS Applications. \emph{Proceedings of 2015 IEEE International Conference on Services Computing (SCC)}, 2015; 210-217.
\bibitem{R32}Alves~A, Arkin~A, Askary~S, Barreto~C, Bloch~B, Curbera~F, Ford~M, Goland~Y, Gu\'{l}zar~A, Kartha~N, Liu~CK, Khalaf~R, K\H{o}nig~D, Marin~M, Mehta~V, Thatte~S, Rijn~D, Yendluri~P, Yiu~A. Web services business process execution language version 2.0. OASIS. 2006.
\bibitem{R33} Papazoglou~MP, Van~Den~Heuvel~WJ. Service oriented architectures: approaches, technologies and research issues. \emph{The VLDB journal} 2007; \textbf{16}(3):389-415.
\bibitem{R34} Vogels~W. Web services are not distributed objects. \emph{IEEE Internet computing} 2003; \textbf{7}(6):59-66.
\bibitem{R35} Baker~S, Dobson~S. Comparing service-oriented and distributed object architectures. \emph{Proceedings of On the Move to Meaningful Internet Systems: OTM} 2005; 631-645
\bibitem{R36}Morin~B, Barais~O, Jezequel~JM, Fleurey~F, Solberg~A. Models@run.time to support dynamic adaptation. \emph{Computer} 2009, \textbf{42}(10):44-51.
\bibitem{R37}Boyer~F, Gruber~O., Pous~D. A robust reconfiguration protocol for the dynamic update of component-based software systems. \emph{Softw. Pract. Exper.} 2017; \textbf{47}(11):1729-1753.
\bibitem{R38}Al-Jaroodi~J, Mohamed~N. Service-oriented middleware: A survey. \emph{Journal of Network and Computer Applications} 2012; \textbf{35}(1):211-220.
\bibitem{R39}Loyall~JP, Gillen~M, Paulos~A, Bunch~L, Carvalho~M, Edmondson~J, Schmidt~DC, Martignoni III~A, Sinclair~A. Dynamic policy-driven quality of service in service-oriented information management systems. \emph{Softw. Pract. Exper.} 2011; \textbf{41}(12):1459-1489.
\bibitem{R40}Lemos~AL, Daniel~F, Benatallah~B. Web service composition: a survey of techniques and tools. \emph{ACM Computing Surveys (CSUR)} 2016; \textbf{48}(3):33.
\bibitem{R41}OMG. Business Process Model Notation (BPMN) version 2.0. OMG Specification. \emph{Object Management Group}, 2011.
\bibitem{R42}Allee~V. Value network analysis and value conversion of tangible and intangible assets. \emph{Journal of Intellectual Capital} 2008; \textbf{9}(1):5-24.
\bibitem{R43} Istio Framework. https://istio.io/. Accessed January 10, 2019
\bibitem{R44} Jamshidi~P, Pahl~C, Mendonca~NC, Lewis~J, Tilkov~S. Microservices: The Journey So Far and Challenges Ahead. \emph{IEEE Software} 2018; \textbf{35}(3):24-35.
\bibitem{R45}Strauch~S, Andrikopoulos~V, S\'{a}ez~SG, Leymann~F. Implementation and Evaluation of a Multi-tenant Open-Source ESB. \emph{Proceedings of European Conference on Service-Oriented and Cloud Computing}, 2013; 79-93.
\bibitem{R46}Blake~M, Wei~T, Rosenberg~F. Composition as a service [web-scale work-flow]. \emph{Internet Computing} 2010; \textbf{14}(1):78-82.
\bibitem{R47}Baresi~L, Guinea~S, Pasquale~L. Service-Oriented Dynamic Software Product Lines with DyBPEL. \emph{Computer} 2012; \textbf{45}(10):42-48.
\bibitem{R48} Tuan~N, Han~J, Colman~A. Modeling and managing variability in process-based service compositions. \emph{Proceedings of the 9th International Conference on Service-Oriented Computing} 2011; 404-420.
\bibitem{R49} Kumara~I, Han~J, Colman~A, Tuan~N, Kapuruge~M. Sharing with a difference: Realizing service-based saas applications with runtime sharing and variation in dynamic software product lines. \emph{Proceedings of 2013 IEEE International Conference on Services Computing} 2013; 567-574.
\bibitem{R50} Eberle~H, Unger~T, Leymann~F. Process fragments. \emph{Proceedings of On the Move to Meaningful Internet Systems: OTM} 2009; 398-405
\bibitem{R51}Walraven~S, Lagaisse~B, Truyen~E, Joosen~W. Policy-driven customization of cross-organizational features in distributed service systems. \emph{Softw. Pract. Exper.} 2013; \textbf{43}(10):1145-1163.
\bibitem{R52} Haesevoets~R, Weyns~D, Holvoet~T. Architecture-centric support for adaptive service collaborations. \emph{ACM Trans. Softw. Eng. Methodol.} 2014; \textbf{23}(1):1-40.
\bibitem{R53} Sherwood~R, Gibb~G, Yap~K, Appenzeller~G, Casado~M, McKeown~N, Parulkar~G. Can the production network be the testbed?. \emph{Proceedings of the 9th USENIX conference on Operating systems design and implementation} 2016; 365-378.
\bibitem{R54} Schmidt~DC. Middleware for real-time and embedded systems. \emph{Communications of the ACM} 2002; \textbf{45}(6):43-48.
\bibitem{R55} Canfora~G, Di~Penta~M, Esposito~R, Perfetto~F, Villani~ML. Service composition (re) binding driven by application-specific qos. \emph{Proceedings of the 4th International Conference on Service-Oriented Computing}, 2006; 141-152.
\bibitem{R56}Van Landuyt~D, Gey~F, Truyen~E, Joosen~W. Middleware for Dynamic Upgrade Activation and Compensations in Multi-tenant SaaS. \emph{Proceedings of International Conference on Service-Oriented Computing}, 2017; 340-348.
\bibitem{R57}Kapuruge~M, Han~J, Colman~A, Kumara~I. ROAD4SaaS: Scalable business service-based saaS applications. \emph{Proceedings of International Conference on Advanced Information Systems Engineering}, 2013; 338-352.
\bibitem{R58}Z\'{u}\~{n}iga-Prieto~M., Gonz\'{a}lez-Huerta~J, Insfran~E, Abrah\r{a}o~S. Dynamic reconfiguration of cloud application architectures. \emph{Softw. Pract. Exper.} 2018; \textbf{48}(2):327-344.
\end{thebibliography}
\end{document}